\documentclass[11pt,a4paper]{article}

\usepackage[margin=1in]{geometry}
\usepackage[T1]{fontenc}
\usepackage[utf8]{inputenc}
\usepackage{lmodern}

\usepackage{amsmath,amssymb,amsthm}
\usepackage{latexsym}
\usepackage{graphicx}
\usepackage{booktabs}
\usepackage{float}
\usepackage{colortbl}
\usepackage{xcolor}
\usepackage{natbib}
\usepackage{hyperref}

\graphicspath{{./images/}}

\setcitestyle{numbers}
\setcitestyle{square}

\colorlet{red}{blue}

\makeatletter
\providecommand{\bmsection}[1]{\section*{#1}}
\renewcommand{\bibsection}{\section*{References}}
\makeatother

\title{Functional Clustering of Survival Data via Smoothed Log-Hazard Trajectories: A Risk-Dynamics Perspective}

\author{
Anna De Magistris\textsuperscript{1}
\quad
Elvira Romano\textsuperscript{1}
\quad
Fabrizio Maturo\textsuperscript{2*}
\\[0.8em]
\small \textsuperscript{1}Department of Mathematics and Physics, University of Campania ``Luigi Vanvitelli'', Caserta, Italy
\\
\small \textsuperscript{2}Department of Economics, Statistics and Business, Universitas Mercatorum, Rome, Italy
\\[0.6em]
\small \textsuperscript{*}Corresponding author: Fabrizio Maturo, \texttt{fabrizio.maturo@unimercatorum.it}
}

\date{}

\begin{document}

\maketitle

\begin{abstract}
\noindent This paper investigates clustering in survival data by shifting the analytical focus from cumulative survival probabilities to instantaneous risk, as characterized by the hazard function. We propose to model smoothed log-hazard trajectories as functional objects that capture the temporal evolution of risk. We propose a clustering framework based on standard Functional Principal Component Analysis (FPCA) applied to B-spline smoothed log-hazard trajectories, which projects these trajectories into a low-dimensional score space while retaining a conservative proportion of their functional variability. The number of retained FPCs is selected before clustering using a 95\% cumulative explained-variance rule, and clustering is then performed on the unstandardized FPCA scores. Clustering is then performed in this space to identify latent group structures in terms of shared risk dynamics. The proposed methodology is evaluated through simulation studies covering progressively complex scenarios, including overlapping and crossing hazard functions. These settings are designed to assess robustness under challenging clinical conditions such as cohort imbalance and heterogeneous risk profiles. We further illustrate the approach on real-world clinical datasets, including the German Breast Cancer Study (GBCS) and the Primary Biliary Cirrhosis (PBC) dataset. Results show that the proposed log-hazard-based functional clustering framework provides an interpretable representation of relative temporal risk dynamics, with competitive internal cohesion and explicit robustness diagnostics when compared with cumulative-survival-based benchmarks.
\end{abstract}


\bigskip

\section{Introduction}\label{sec.1}

In the current era of growing reliance on data-driven insights for informed decision-making, survival analysis stands out by providing valuable information about the timing and risk of events. Its applications now extend beyond traditional fields such as medicine and epidemiology to areas like engineering, marketing, sociology, and many others. Clustering is one of the many statistical methods used in survival analysis and can play a crucial role in uncovering hidden structures within data. 

However, at the core of this work lies an essential distinction. When referring to survival clustering, the focus is typically on what we define here as \textit{individual-level clustering}, which involves clustering individuals based on their survival times and associated covariates, allowing researchers to identify detailed patterns and heterogeneity within the population. This means that clustering focuses solely on the raw survival data and other related variables, aiming to reveal subgroups of individuals with distinct survival patterns. Recent state of the art approaches in this domain have increasingly leveraged deep learning frameworks \citep{mouli2019deep, manduchi2021deep} and advanced mixture models \citep{buginga2024clustering, chapfuwa2020survival} to integrate latent representation learning directly with survival prediction. This individual-level method has been extensively studied and applied, and remains a cornerstone of survival analysis literature. Nevertheless, another perspective on survival clustering is largely underexplored, despite its high potential utility. This alternative approach, which we define as \textit{curve-level clustering}, focuses on grouping survival curves, such as Kaplan-Meier curves, rather than individual survival times. By clustering curves that represent aggregated data from multiple cohorts or groups, this method uncovers broader trends that might be overlooked when analyzing individual data alone. 

\begin{figure*}[htbp] 
\centering
\begin{minipage}[t]{0.49\textwidth}
\centering
\includegraphics[width=\textwidth]{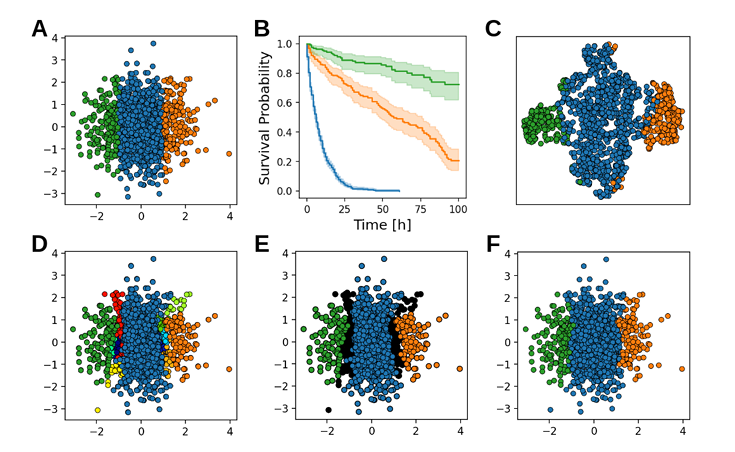} 
\end{minipage}
\hfill
\begin{minipage}[t]{0.49\textwidth}
\centering
\includegraphics[width=\textwidth]{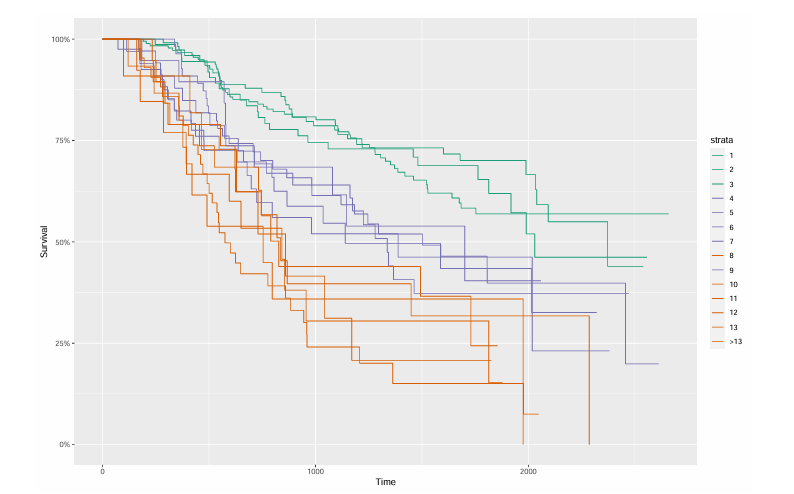} %
\end{minipage}
\caption{\textit{On the left, a visualization of the individual-level clustering process in \citep{10.1007/978-3-031-34344-5_1}; on the right, an example of curve-level clustering outcome in \citep{RJ-2021-032}.}}
\label{fig:clustering_comparison}
\end{figure*}

Curve-level clustering is particularly valuable in contexts where aggregated survival outcomes provide clearer insights. In multicenter studies, for instance, it helps compare survival across hospitals following a common treatment protocol, identifying patterns that can be used to enhance patient care. In epidemiology, it might highlight disparities in disease incidence or mortality across regions, countries or demographic aspects, informing public health strategies, such as resource allocation and intervention planning. For instance, functional clustering methods applied to age-specific mortality curves have successfully revealed distinct patterns and stages of mortality decline across different nations \citep{leger2021functional}. Moreover, the approach can easily apply to reliability analysis in engineering, comparing failure rates of mechanical systems or manufactured products across factories, as well as to marketing, where it aids in customer churn analysis, or sociology, where it supports event history analysis by studying patterns in events such as job changes or criminal recidivism. 

The broad applicability of curve-level clustering underscores the need for further research. To the best of our knowledge, the only method currently dedicated to clustering survival curves is \textit{Survclustcurves} \citep{RJ-2021-032}, which clusters Kaplan-Meier curves using a log-rank-based dissimilarity matrix and subsequently K-means or K-medoids algorithms. Recent literature has also sought to improve the computational efficiency of grouping multiple survival curves by combining K-means directly with log-rank tests to avoid expensive bootstrap approximations \citep{villanueva2025efficient}. These approaches provide important benchmarks for curve-level survival clustering, but they primarily operate on cumulative survival information. As a result, they may be less sensitive to differences in the timing and localization of instantaneous risk, especially when survival curves overlap, cross, or converge over the follow-up period.

While benchmark methods essentially operate on discrete non-parametric step functions, this paper proposes a Functional Data Analysis (FDA) \citep{ramsay2005functional, gertheiss2023functional} approach. FDA-based representations have already shown their usefulness in biomedical settings where the relevant information is encoded in the shape of complex temporal signals rather than in isolated scalar summaries, including supervised classification of electrocardiographic signals \citep{MaturoVerde2022}. A recent contribution more directly related to survival modeling is provided by \citet{Loffredo2025}, who developed a Random Survival Forest framework for censored functional data, showing how functional representations can be integrated into survival analysis beyond classical scalar covariates. The adoption of FDA in the present work is not merely justified by the assumption that survival processes are inherently continuous. Rather, it is driven by the premise that the most relevant prognostic information is embedded in the global structure of the trajectories along the time domain. Differences between groups rarely manifest solely as distinct pointwise values; they emerge in the overall shape, the timing and magnitude of risk peaks, and the temporal dynamics of the process. Although the curves we analyze are not direct functional observations, but rather discrete clinical estimates, mapping them into a functional form allows us to explicitly model the dependence between adjacent time points, treat each curve as a single cohesive unit, and decompose their between-curve variability into interpretable components. In the present work, this functional representation is constructed on the log-hazard scale after B-spline smoothing, so that the object analyzed by FPCA is the smoothed log-hazard trajectory rather than the raw empirical hazard estimate.

Our methodology shifts the analytical focus from cumulative survival probabilities to the instantaneous hazard function. This choice is not motivated by mathematical convenience, but by the richer informational content of the hazard. The hazard function describes the instantaneous dynamics of risk over time, allowing us to identify and distinguish trajectories that might appear indistinguishable or heavily obscured when aggregated into cumulative survival curves. Indeed, cohorts with similar final survival outcomes can exhibit profoundly different underlying risk behaviors. Once the hazard is selected to capture these fine temporal dynamics, FDA naturally emerges as the most appropriate tool to represent and analyze its continuous structure. As an added, though secondary, advantage, the fact that the hazard function is not constrained to be monotonically decreasing makes its functional representation more flexible and less mathematically constrained than cumulative curves. Because empirical hazard estimates can be locally unstable, especially in intervals with sparse events or limited numbers at risk, the analysis is performed on smoothed log-hazard trajectories. This choice preserves positivity after back-transformation and provides a regularized functional object suitable for FPCA and clustering.

By representing smoothed log-hazard trajectories within this functional framework, the objective of clustering translates into identifying groups of curves that share similar patterns of risk variation over time. To achieve this, we apply standard Functional Principal Component Analysis (FPCA). Here, FPCA is not utilized merely as a dimensionality reduction technique, but as a method to construct a space where differences between curves are explicitly expressed in terms of principal modes of variation. The number of retained FPCs is selected before clustering using a conservative cumulative explained-variance criterion, retaining the minimum number of components required to reach or exceed 95\% of the total functional variability. Consequently, performing clustering on the resulting unstandardized functional scores carries a precise statistical meaning: we are aggregating trajectories that share similar dominant patterns in the temporal dynamics of risk, rather than trajectories that are necessarily identical in every local feature. 

Within this FPCA space, we explore the data structure using various clustering algorithms, including K-means, K-medoids, and Hierarchical clustering with Ward's minimum variance linkage. Rather than focusing primarily on the specific properties of these algorithms, we employ them as alternative tools to uncover latent structures. The number of clusters is selected after the FPCA representation has been fixed, using internal validity criteria and stability diagnostics. This separation avoids choosing the functional representation solely on the basis of apparent clustering performance and helps preserve relevant temporal variability before the clustering step.

The proposed framework is evaluated through three complementary levels of analysis. First, subject-level simulations are used to assess the ability of the method to recover known latent risk regimes under increasing structural complexity and outlier contamination. Second, real-data applications are used to illustrate the behavior of the method in observational clinical settings where no ground-truth partition is available. Third, sensitivity analyses are performed by varying the FPCA explained-variance threshold from 90\% to 99\%, in order to assess whether the identified groups remain stable when the amount of retained functional variability changes.

The remainder of this paper is organized as follows. Section \ref{sec.2} details the mathematical foundation of our proposed functional methodology. Section \ref{sec.results} evaluates the framework through simulation studies under varying degrees of complexity, including imbalanced cohorts. Section \ref{sec.real_data} demonstrates the applicability of our approach on two real-world clinical datasets. Finally, Section \ref{sec:con} provides concluding remarks and outlines future research directions.

\section{Materials and Methods}\label{sec.2}
\subsection{A functional representation of log-hazard trajectories}

Let $T$ be a non-negative random variable representing the time-to-event, with probability density function $f(t)$ and survival function defined as
\[
S(t) = \Pr(T > t).
\]

The hazard function is defined as the instantaneous event rate conditional on survival up to time $t$:
\[
h(t) = \lim_{\Delta t \to 0} \frac{\Pr(t \le T < t+\Delta t \mid T \ge t)}{\Delta t}.
\]

Hazard functions are not observed directly but are obtained through transformations of estimated survival or density quantities. As a consequence, empirical hazard estimates are typically highly variable and sensitive to sampling noise. In particular, operations such as numerical differentiation (e.g., $h(t) = -\frac{d}{dt}\log S(t)$) or indirect reconstruction from discrete survival curves tend to amplify irregularities in the data, producing unstable trajectories.

Raw estimators such as Kaplan--Meier curves or plug-in hazard estimates cannot be directly used within a functional framework. Without smoothing, these operations remain mathematically well defined for step functions, but they may become statistically unstable and highly sensitive to the chosen discretization, sparse events, and local estimation noise.
\[
\int_{\tau} \hat{h}_i(t)\,\phi(t)\,dt
\]
and $L^2(\tau)$ distances
\[
\|\hat h_i - \hat h_j\|_2^2 = \int_{\tau} \big(\hat h_i(t) - \hat h_j(t)\big)^2 dt
\]
become highly unstable when evaluated on step functions or noisy discrete approximations.

Smoothing therefore restores continuity and square-integrability and by projecting discrete estimators onto a smooth functional space, we obtain stable and differentiable trajectories that are compatible with the geometry of $L^2(\tau)$.
For this reason, we treat the estimated hazard functions as realizations of a stochastic process in $L^2(\tau)$, where $\tau \subset \mathbb{R}^+$ is a compact time interval. This space is equipped with the standard inner product
\[
\langle f,g\rangle = \int_{\tau} f(t)g(t)\,dt.
\]

\medskip
At this point, it is important to clarify why the functional representation is constructed on the hazard function rather than directly on the survival function. Although both quantities originate from the same probabilistic structure, they exhibit different geometric properties when viewed as elements of a functional space.

The survival function $S(t)$ is intrinsically constrained: it is monotone non-increasing, bounded in $[0,1]$, and its variability across cohorts is largely compressed by its cumulative nature. Structural differences between groups are often absorbed into the overall decline of the curve, making distinct temporal risk patterns appear geometrically similar once integrated over time. As a consequence, the space of admissible survival curves is highly restricted, and functional variability is limited to a narrow set of shapes.

In contrast, the hazard function $h(t)$ is not subject to monotonicity or boundedness constraints. It can exhibit peaks, inflection points, temporal accelerations, and localized variations that directly reflect the instantaneous dynamics of risk. From a functional perspective, this makes the hazard a far richer object: the principal modes of variation between cohorts are expressed in the shape and timing of these features, rather than being flattened by cumulative aggregation.

This distinction is crucial for Functional Principal Component Analysis. FPCA aims to identify dominant modes of functional variation, and such modes can only emerge clearly in a space where curves are free to express structural differences. While survival curves tend to concentrate variability along a single dominant direction corresponding to overall decline, hazard functions distribute variability across multiple temporal patterns. For this reason, modeling $h(t)$ provides a geometrically expressive space in which FPCA can effectively capture interpretable differences between trajectories. Operationally, however, the FPCA is applied to the corresponding smoothed log-hazard trajectories, which provide a stable unconstrained representation while preserving positivity after back-transformation.

To ensure numerical stability and to guarantee that the estimated risk remains strictly positive and consistent with probability theory ($h(t) \ge 0$), we perform the functional representation on the logarithmic scale. Let $\hat h_i(t)$ be the raw, discrete hazard estimate for the $i$-th cohort. To avoid numerical instability in regions with zero or nearly zero empirical hazard, a small positive floor is applied before the logarithmic transformation. We define the smoothed log-hazard trajectory via a continuous B-spline basis expansion:
\[
\log \hat h_i(t) = \sum_{k=1}^{n_\beta} c_{i,k} B_k(t),
\]
where $\{B_k(t)\}_{k=1}^{n_\beta}$ denotes a set of B-spline basis functions evaluated on the time grid, and $\mathbf{c}_i = (c_{i,1},\dots,c_{i,n_\beta})$ are the regularized smoothing coefficients.
It is important to emphasize that FPCA is applied to these B-spline smoothed log-hazard curves, meaning it analyzes the variability of the regularized functions rather than the noisy raw data. The level of smoothing directly influences the amount and nature of the variability captured: a more rigid representation may attenuate local patterns, whereas a more flexible one might emphasize short-term fluctuations, with subsequent effects on the interpretation of the principal components and the resulting cluster structure. For this reason, the B-spline smoothing parameters are selected through a GCV-constrained flexibility rule rather than by mechanically choosing only the absolute minimum of the GCV curve. Let \(\mathrm{GCV}_{\min}\) denote the minimum mean GCV value over the candidate B-spline configurations. We first define the admissible set
\[
\mathcal{A}_{\mathrm{GCV}}(\delta)
=
\left\{
\theta:
\mathrm{GCV}(\theta)
\leq
(1+\delta)\,\mathrm{GCV}_{\min}
\right\},
\qquad
\delta \geq 0.
\]
In the empirical analyses, the relative GCV-equivalence tolerance was fixed at \(\delta=0.20\). Candidate smoothers were evaluated over the B-spline basis grid
\[
n_\beta \in \{10,12,15,20,25,30,35,40,50,60,70,80\}
\]
and over a logarithmic smoothing-penalty grid \(\lambda=10^a\), with 60 equally spaced values of \(a\) between \(-18\) and \(-1\). After identifying the candidate with minimum mean GCV, all configurations whose mean GCV was within 20\% of this minimum were considered GCV-equivalent. Among these admissible candidates, the final smoother was selected by maximizing a local-variability score, defined as the average integrated squared second derivative of the smoothed log-hazard trajectories. Ties were then resolved by favoring larger mean effective degrees of freedom, a larger number of B-spline basis functions, a smaller smoothing penalty, and finally a smaller mean GCV. This criterion preserves GCV as the primary regularization criterion, while allowing sufficient flexibility to retain interpretable local features of the log-hazard dynamics. This choice is particularly relevant in the present setting, because hazard functions may contain localized temporal features, such as short-term accelerations, delayed peaks, or transient risk increases, that can be attenuated by an excessively rigid smoothing rule. The adopted criterion therefore balances statistical regularization with the need to preserve interpretable local variability in the risk dynamics. This choice is particularly relevant in the present setting, because hazard functions may contain localized temporal features, such as short-term accelerations, delayed peaks, or transient risk increases, that can be attenuated by an excessively rigid smoothing rule. The adopted criterion therefore balances statistical regularization with the need to preserve interpretable local variability in the risk dynamics.

Because the analysis is performed on the natural logarithm of the hazard, no restrictive non-negativity constraints must be imposed on the basis coefficients $c_{i,k}$. The final reconstructed empirical hazard curve is simply obtained via exponentiation:
\[
\hat h_i(t) = \exp \left( \sum_{k=1}^{n_\beta} c_{i,k} B_k(t) \right) \ge 0,
\]
which inherently satisfies the strict positivity requirement for all $t \in \tau$.

The log-hazard scale also has an important interpretative implication. Differences between two log-hazard trajectories correspond to logarithms of hazard ratios:
\[
\log h_i(t)-\log h_j(t)=\log\left\{\frac{h_i(t)}{h_j(t)}\right\}.
\]
Therefore, proximity between trajectories on the log-hazard scale reflects similarity in relative, rather than purely absolute, risk dynamics. In this sense, the functional geometry used by FPCA is naturally aligned with a multiplicative interpretation of risk, since a constant vertical difference on the log-hazard scale corresponds to a constant hazard ratio on the original scale.

Functional Principal Component Analysis \citep{ramsay2005functional} is then applied to these regularized log-hazard trajectories. FPCA effectively projects the functions onto a low-dimensional orthogonal basis, capturing their structural variations.

Let
\[
\mu(t) = \frac{1}{N}\sum_{i=1}^N \log \hat h_i(t)
\]
be the mean log-hazard function. FPCA is defined by solving
\[
\max_{\phi} \ \mathrm{Var}\!\left(\int_{\tau}\log \hat h_i(t)\phi(t)\,dt\right)
\quad \text{subject to} \quad \|\phi\|_2 = 1,
\]
yielding eigenfunctions $\{\phi_m(t)\}_{m \ge 1}$ and associated scores
\[
\xi_{im} = \int_{\tau} \big(\log \hat h_i(t) - \mu(t)\big)\phi_m(t)\,dt.
\]

According to the Karhunen--Loève theorem \citep{karhunen1947}, each estimated log-hazard trajectory admits the approximation
\[
\log \hat h_i(t) \approx \mu(t) + \sum_{m=1}^{M} \xi_{im}\phi_m(t),
\]
which successfully maps the functional observations into a finite-dimensional Euclidean representation in $\mathbb{R}^{M}$ for downstream clustering. Consequently, all subsequent clustering operations are performed on the FPCA scores of the smoothed log-hazard trajectories. Distances in the score space should therefore be interpreted as approximations of functional dissimilarities between log-hazard trajectories, rather than between hazards on the original scale.

\subsection{Selection of the number of retained functional principal components}

A key methodological choice in the proposed framework concerns the truncation level of the Functional Principal Component Analysis expansion, namely the number \(M\) of functional principal components retained before clustering. In this study, \(M\) is not selected by directly maximizing a clustering performance index. Instead, it is selected before clustering according to a conservative cumulative explained-variance criterion, whose purpose is to preserve the functional information contained in the smoothed log-hazard trajectories.

Let \(\lambda_1 \geq \lambda_2 \geq \cdots \geq \lambda_q\) denote the eigenvalues associated with the functional principal components extracted from the smoothed log-hazard trajectories. The cumulative proportion of functional variance explained by the first \(M\) components is defined as
\[
\mathrm{CPV}(M)
=
\frac{\sum_{m=1}^{M} \lambda_m}
{\sum_{m=1}^{q} \lambda_m}.
\]

The number of retained components is selected as
\[
M^{\ast}
=
\min
\left\{
M:
\mathrm{CPV}(M) \geq 0.95
\right\}.
\]
This rule means that we retain the smallest number of FPCs required to reach or exceed 95\% of the total functional variance. Since the cumulative explained variance increases in discrete steps, the selected value does not generally explain exactly 95\%, but is the first truncation level whose cumulative variance is at least 95\%.

Clustering is then performed on the unstandardized FPCA score vectors
\[
\boldsymbol{\xi}_i
=
(\xi_{i1},\ldots,\xi_{iM^{\ast}}),
\qquad
i=1,\ldots,N,
\]
where \(\xi_{im}\) is the score of the \(i\)-th smoothed log-hazard trajectory on the \(m\)-th functional principal component. The use of unstandardized scores is important because it preserves the variance structure induced by the FPCA eigenvalues. Standardizing the scores before clustering would assign the same weight to leading and secondary components, potentially giving excessive influence to low-variance directions that may reflect residual noise or minor local fluctuations. Standardized scores, when used, are therefore restricted to graphical diagnostics and not to clustering, cluster selection, or stability assessment.

The use of a 95\% cumulative explained-variance threshold is motivated by the functional nature of the data and by the objective of the analysis. The aim is not merely to obtain a low-dimensional representation with strong apparent cluster separation, but to perform clustering on a representation that retains the main temporal structure of the smoothed log-hazard trajectories. These trajectories may differ not only in their overall level, but also in the timing, intensity, persistence, and localization of risk peaks. Such features are central to the interpretation of dynamic risk patterns. An overly aggressive dimensionality reduction, for example retaining only components explaining 70\% or 80\% of the variability, could remove local or medium-scale temporal differences that are substantively relevant for the analysis of hazard dynamics.

For this reason, the 95\% threshold is used as a conservative compromise. It retains most of the systematic functional variability in the smoothed log-hazard curves, while discarding only a small residual fraction of variability. This residual part is expected to contain mainly minor oscillations, estimation noise, or highly local fluctuations that are less likely to reflect stable and interpretable risk dynamics. In this sense, the FPCA truncation is not intended to optimize the clustering result ex post, but to ensure that clustering is performed on a representation that remains faithful to the functional information carried by the log-hazard trajectories.

This choice also separates the role of dimensionality reduction from the role of cluster validation. The cumulative explained-variance criterion determines the functional representation used as input for clustering. The number of clusters \(K\) is then selected only after the FPCA score space has been fixed, using internal validity criteria such as the average Silhouette width and, where applicable, stability diagnostics. In simulated scenarios, where the true data-generating groups are known, the Adjusted Rand Index is used only as an external recovery measure. In real-data applications, where no ground-truth partition is available, internal measures and stability diagnostics are used to assess cohesion, separation, and robustness of the identified groups.

To further assess the impact of the FPCA truncation rule, the empirical applications also include a sensitivity analysis in which the explained-variance threshold is varied over a continuum from 90\% to 99\%. For each threshold value \(c \in [0.90,0.99]\), the number of retained components is defined as
\[
M(c)
=
\min
\left\{
M:
\mathrm{CPV}(M) \geq c
\right\},
\]
and the clustering procedure is repeated using the corresponding unstandardized FPCA score vectors
\[
\boldsymbol{\xi}_i(c)
=
(\xi_{i1},\ldots,\xi_{iM(c)}).
\]

The partition obtained under the main 95\% rule is used as the reference solution. The partitions obtained under alternative thresholds are compared with this reference partition in terms of cluster composition, internal validity, and agreement measures. This sensitivity analysis is reported in the application section and is used to verify whether the main empirical conclusions remain stable when the amount of retained functional variability is varied. A stable clustering structure across the interval from 90\% to 99\% indicates that the identified groups are not an artifact of a specific truncation threshold, but reflect persistent patterns in the temporal dynamics of the smoothed log-hazard trajectories.

\subsection{Clustering log-hazard dynamics in the FPCA score space}

The core methodological contribution of this work lies in the clustering of smoothed log-hazard trajectories in a functional setting. Starting from discrete survival data, we construct smooth log-hazard functions and represent them as functional objects that capture the temporal dynamics of the underlying risk process. The goal is to identify latent groups of subjects based on similarities in their log-hazard evolution over time.
Operationally, the proposed approach follows a two-stage clustering strategy as defined by \citep{jacques2014functional}. In the first stage, the functional log-hazard trajectories are mapped onto a finite-dimensional representation that preserves their main modes of variation. In the second stage, standard clustering algorithms are applied in this reduced space to identify homogeneous groups of temporal dynamics. The dimensionality of this score space is fixed before clustering through the cumulative explained-variance rule described above, so that the choice of the functional representation is separated from the subsequent selection of the number of clusters.

Let $\boldsymbol{\xi}_i \in \mathbb{R}^M$ denote the FPCA score vector associated with the $i$-th smoothed log-hazard trajectory. The collection
\[
\mathcal{X}_M = \{\boldsymbol{\xi}_1,\dots,\boldsymbol{\xi}_N\}
\]
is endowed with the standard Euclidean metric
\[
d_M(x,y) = \|x-y\|_2,
\]
which provides a representation of the functional variability. All clustering procedures are applied to the unstandardized FPCA scores. This preserves the variance structure induced by the FPCA eigenvalues, whereas standardizing the scores before clustering would give the same operational weight to leading and low-variance components. Standardized scores are used only for graphical diagnostics, when component-wise score distributions need to be visually compared.

Clustering is then defined as a partition
\[
\mathcal{P}_K = \{\mathcal{C}_1,\dots,\mathcal{C}_K\}, \qquad \bigcup_{k=1}^K \mathcal{C}_k = \{1,\dots,N\},
\]
obtained by minimizing a within-cluster distortion functional of the form
\[
\mathcal{L}(\mathcal{P}_K)
=
\sum_{k=1}^K \sum_{i \in \mathcal{C}_k}
\ell\!\left(\boldsymbol{\xi}_i, \boldsymbol{\theta}_k\right),
\]
where $\boldsymbol{\theta}_k$ denotes a cluster representative and $\ell(\cdot,\cdot)$ is a dissimilarity measure in $\mathbb{R}^M$.

In the case of K-means clustering, $\ell(x,\theta)=\|x-\theta\|_2^2$ with
\[
\boldsymbol{\theta}_k = \frac{1}{|\mathcal{C}_k|}\sum_{i \in \mathcal{C}_k} \boldsymbol{\xi}_i,
\]
leading to the standard optimization problem
\[
\min_{\mathcal{P}_K} \sum_{k=1}^K \sum_{i \in \mathcal{C}_k} \|\boldsymbol{\xi}_i - \boldsymbol{\theta}_k\|_2^2.
\]

For K-medoids, representatives are restricted to the empirical support,
\[
\boldsymbol{\theta}_k \in \{\boldsymbol{\xi}_1,\dots,\boldsymbol{\xi}_N\},
\]
and the objective becomes
\[
\min_{\mathcal{P}_K} \sum_{k=1}^K \sum_{i \in \mathcal{C}_k} \|\boldsymbol{\xi}_i - \boldsymbol{\theta}_k\|_2.
\]

Hierarchical clustering is defined through an agglomerative sequence of partitions
\[
\mathcal{P}_N \to \mathcal{P}_{N-1} \to \cdots \to \mathcal{P}_1,
\]
where at each step two clusters are merged according to a linkage criterion. In the Ward formulation, the merge is selected as
\[
(\mathcal{C}_a,\mathcal{C}_b)
=
\arg\min_{a \neq b}
\Delta(\mathcal{C}_a,\mathcal{C}_b),
\]
with $\Delta$ denoting the increase in within-cluster variance induced by the merge.

The FPCA-based representation ensures that clustering is performed on the dominant modes of variation of the smoothed log-hazard trajectories, rather than on pointwise or noisy features of the original functional data. In particular, the truncated Karhunen--Loève expansion provides a finite-dimensional approximation that captures the essential structure of the log-hazard dynamics, making Euclidean distances between score vectors a stable proxy for functional dissimilarity in $L^2(\tau)$. Since the FPCA is applied on the log-hazard scale, these distances approximate dissimilarities between log-hazard trajectories, not between hazards on the original scale. Consequently, the resulting clusters should be interpreted as groups with similar relative temporal risk dynamics. More precisely, a difference between two log-hazard trajectories corresponds pointwise to the logarithm of a hazard ratio. Thus, clustering in the FPCA score space tends to group trajectories according to multiplicative patterns of instantaneous risk over time, rather than according to equal absolute differences in the hazard scale.

A critical component of the clustering framework is the determination of the optimal number of clusters $K$. Because survival clustering often aims to uncover latent structures without a known ground truth, a robust, data-driven selection criterion is essential. We formalize this selection by evaluating the geometric cohesion and separation of candidate partitions using the Silhouette index \citep{rousseeuw1987silhouettes}. 

For a given partition $\mathcal{P}_K$ and an observation $\boldsymbol{\xi}_i$ assigned to cluster $\mathcal{C}_k$, let $a(\boldsymbol{\xi}_i)$ be the average Euclidean distance between $\boldsymbol{\xi}_i$ and all other points in $\mathcal{C}_k$, representing intra-cluster cohesion:
\[
a(\boldsymbol{\xi}_i) = \frac{1}{|\mathcal{C}_k| - 1} \sum_{\boldsymbol{\xi}_j \in \mathcal{C}_k, j \neq i} \|\boldsymbol{\xi}_i - \boldsymbol{\xi}_j\|_2.
\]
Let $b(\boldsymbol{\xi}_i)$ be the minimum average distance from $\boldsymbol{\xi}_i$ to points in any other cluster $\mathcal{C}_l$ ($l \neq k$), representing inter-cluster separation:
\[
b(\boldsymbol{\xi}_i) = \min_{l \neq k} \frac{1}{|\mathcal{C}_l|} \sum_{\boldsymbol{\xi}_j \in \mathcal{C}_l} \|\boldsymbol{\xi}_i - \boldsymbol{\xi}_j\|_2.
\]
The Silhouette value for the $i$-th observation is defined as:
\[
s(\boldsymbol{\xi}_i) = \frac{b(\boldsymbol{\xi}_i) - a(\boldsymbol{\xi}_i)}{\max\{a(\boldsymbol{\xi}_i), b(\boldsymbol{\xi}_i)\}}.
\]
The global clustering quality for a partition $\mathcal{P}_K$ is quantified by the average Silhouette width:
\[
\bar{S}(K) = \frac{1}{N} \sum_{i=1}^N s(\boldsymbol{\xi}_i).
\]
To select the optimal $K$, denoted as $K_{opt}$, we evaluate a set of candidate values $\mathcal{K} = \{2, \dots, K_{max}\}$. For each candidate algorithm and each candidate value of \(K\), the average Silhouette width is computed in the selected unstandardized FPCA score space. Rather than treating negligible numerical differences in Silhouette as substantively meaningful, candidate solutions whose Silhouette values lie within a small tolerance of the maximum are regarded as practically equivalent. Among these near-equivalent solutions, a parsimonious tie-breaking rule is applied, favoring the smaller number of clusters and then the predefined order of clustering methods. Thus, the final choice of \(K\) is not a blind maximization of the Silhouette index, but a parsimonious selection among solutions with comparable internal cohesion and separation.

In simulation studies, where the true latent partition is known by construction, external recovery measures such as the Adjusted Rand Index are used to evaluate the recovered partition. In real-data applications, where no ground-truth partition is available, the assessment relies on internal validity, stability diagnostics, and functional interpretability. Therefore, ARI is used as a recovery metric in simulated settings, not as a general criterion for selecting the number of retained FPCs or the number of clusters in observational applications.

\section{Performance evaluation over a continuum of complexity}\label{sec.results}

To evaluate the robustness of the proposed framework, we constructed a simulation environment characterized by a \textit{continuum} of increasing structural complexity. Rather than relying on rigid parametric distributions, this continuum was achieved by progressively modulating the separation between the latent temporal risk regimes. Specifically, we linearly interpolated the B-spline coefficient matrices defining the early, intermediate, and late log-hazard peaks toward a single, common global mean trajectory. 

The simulation progresses from a baseline state ($0\%$ overlap), where the three clusters exhibit highly distinct functional profiles, up to a confounding state ($50\%$ overlap). By limiting the continuum to this threshold, we observe the algorithms' behavior through a realistic degradation of the signal-to-noise ratio, just prior to the point where the functional profiles become completely indistinguishable. For each discrete $5\%$ increment of overlap, we evaluated the ability of the algorithms to correctly recover the ground truth assignment.

\begin{figure*}[!t]
    \centering
    \includegraphics[width=0.75\textwidth]{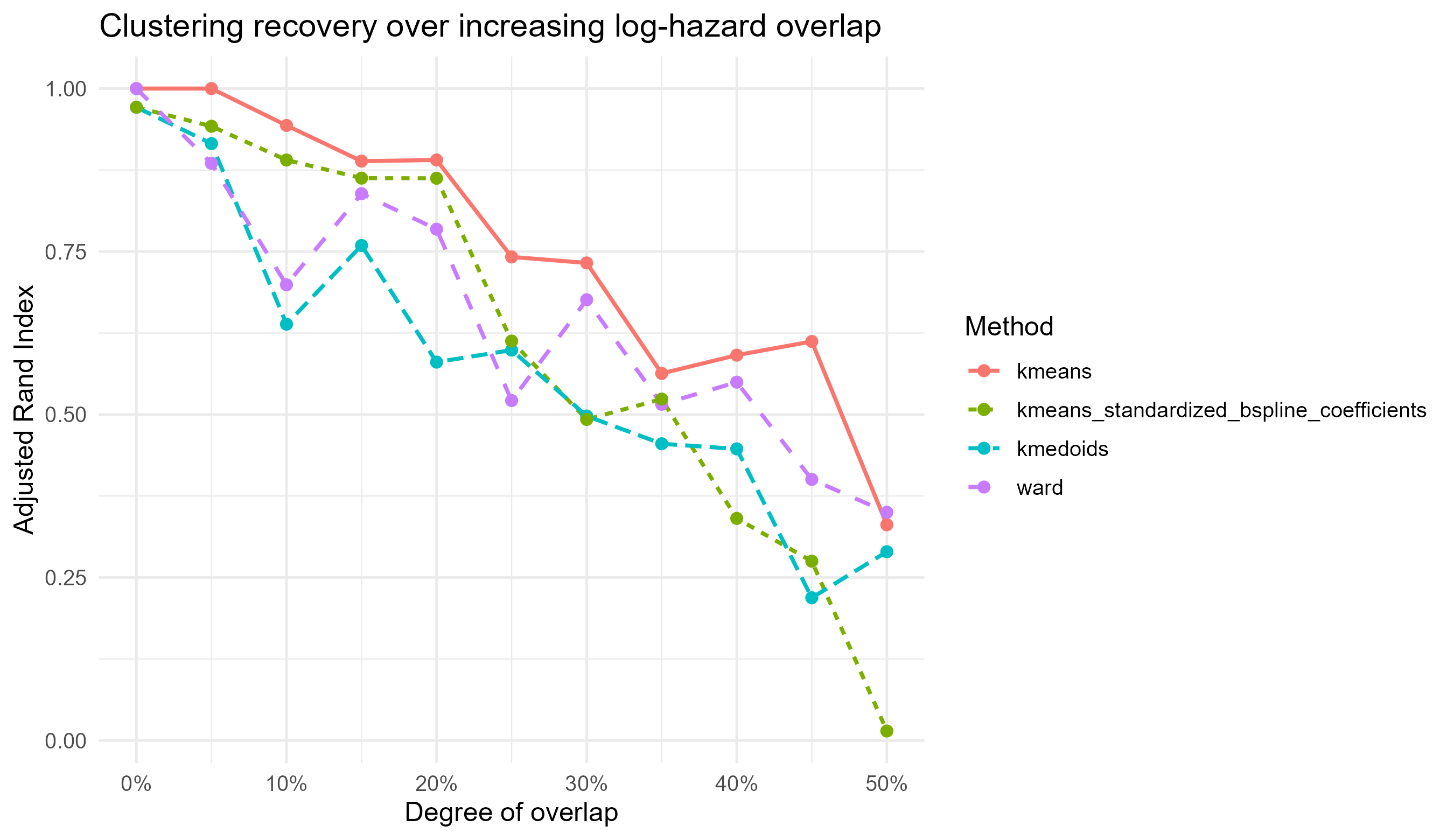}
    \caption{Clustering recovery over increasing log-hazard overlap. The figure reports the mean Adjusted Rand Index for recovering the true \(K=3\) latent groups as the degree of overlap between the simulated log-hazard regimes increases from \(0\%\) to \(50\%\). The comparison includes the proposed FPCA-based clustering strategies, namely K-means, K-medoids, and Ward clustering on unstandardized FPCA scores, and a benchmark based on K-means applied to standardized B-spline coefficients.}
    \label{fig:continuum_plot}
\end{figure*}

As illustrated in Figure \ref{fig:continuum_plot}, all evaluated methods successfully identify the latent structure under near-ideal separation, with ARI values close to 1 at $0\%$ overlap. As the degree of overlap increases, however, the recovery performance of all methods deteriorates, as expected, revealing marked differences in robustness across clustering strategies.

Among the evaluated procedures, K-means applied to the FPCA score space consistently provides the strongest overall recovery profile. Its ARI remains the highest, or among the highest, across most of the overlap continuum, and degrades more gradually than the alternative methods. In particular, it maintains ARI values around or above 0.60 up to approximately $45\%$ overlap, indicating that the low-dimensional FPCA representation preserves the dominant group-specific temporal risk patterns even under substantial confounding.

Ward hierarchical clustering also shows competitive behavior, especially at moderate-to-high overlap levels, and remains comparatively stable in the most challenging part of the continuum. Functional K-medoids exhibits a broadly similar declining pattern, but with somewhat greater sensitivity in the intermediate overlap region. Both methods remain substantially informative across a large portion of the continuum, although they are generally less stable than K-means in the FPCA space.

The benchmark based on K-means clustering of standardized B-spline coefficients performs reasonably well when the latent groups are well separated, but its degradation becomes more pronounced in the most difficult scenarios. In particular, after approximately $40\%$ overlap, its recovery drops sharply and approaches zero at $50\%$ overlap. This suggests that clustering directly on standardized spline-coefficient representations may become unstable when local temporal differences are heavily blurred by overlap, whereas the FPCA representation offers a more robust summary of the main log-hazard dynamics.

The continuum experiment supports the methodological role of the proposed FDA-based pipeline. The advantage of the approach does not lie merely in smoothing the trajectories, but in combining a regularized log-hazard representation with FPCA-based dimensionality reduction before clustering. This yields a more gradual loss of recovery as the latent risk regimes become increasingly similar, and therefore provides a more robust strategy for identifying dynamic survival patterns under realistic structural complexity.

\subsection{Sensitivity analysis with respect to the FPCA truncation threshold in the simulation study}

To assess whether the recovery performance in the simulation study depends on the specific 95\% FPCA truncation rule, we repeated the clustering analysis over a grid of cumulative explained-variance thresholds ranging from 90\% to 99\%. For each threshold \(c\), the number of retained components was defined as
\[
M(c)
=
\min\left\{
M:
\mathrm{CPV}(M)\geq c
\right\}.
\]
The clustering procedures were then re-estimated using the corresponding unstandardized FPCA score vectors, and the resulting partitions were compared with the known latent groups through the Adjusted Rand Index.

\begin{figure*}[!t]
    \centering
    \includegraphics[width=0.78\textwidth]{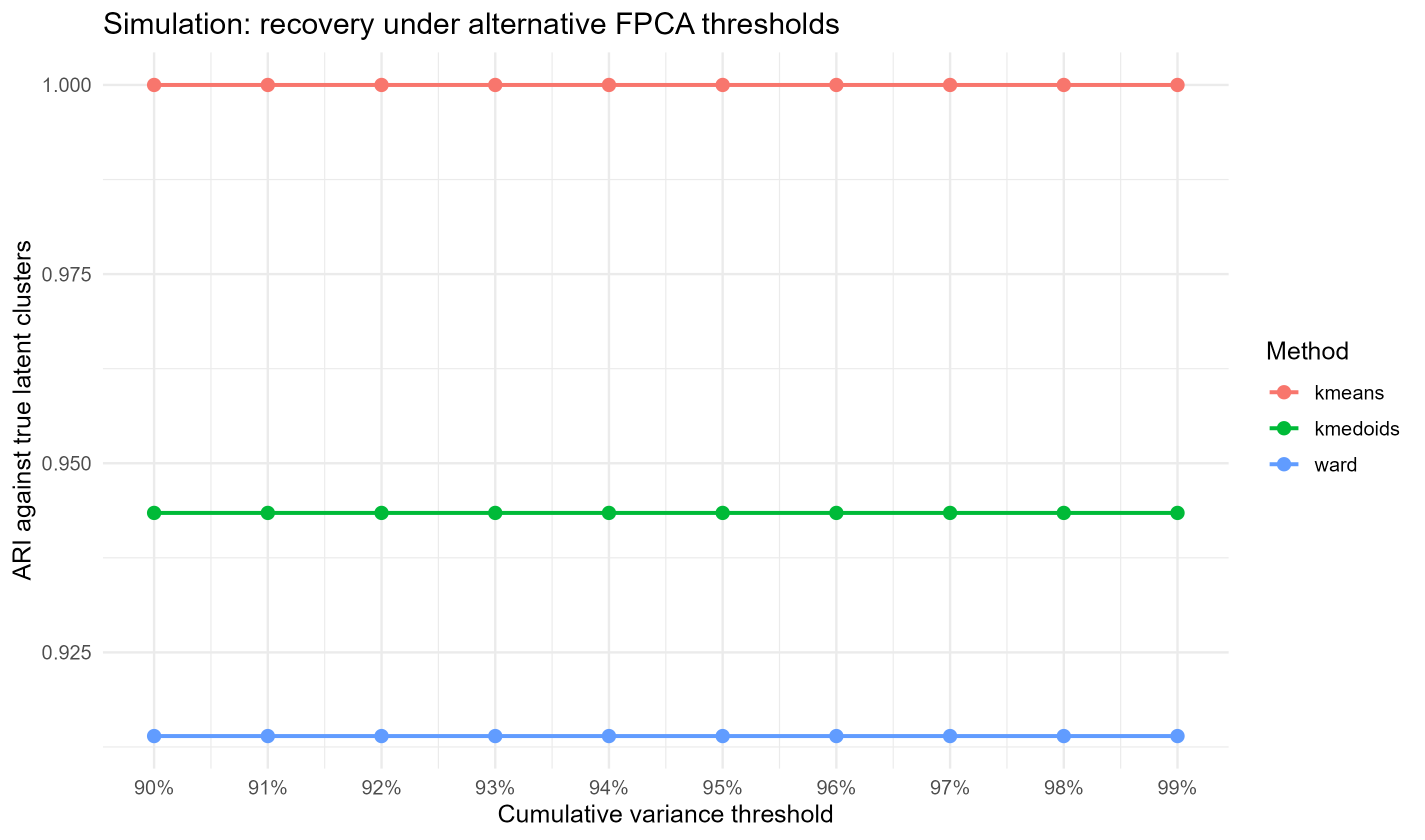}
    \caption{Simulation sensitivity analysis under alternative FPCA truncation thresholds. The figure reports the Adjusted Rand Index against the true latent partition for thresholds ranging from 90\% to 99\%.}
    \label{fig:simulation_threshold_sensitivity}
\end{figure*}

Figure~\ref{fig:simulation_threshold_sensitivity} shows that the recovery performance is essentially invariant across the entire range of FPCA truncation thresholds. K-means on FPCA scores achieves perfect recovery for all thresholds, with ARI equal to 1.000. K-medoids and Ward clustering also show constant recovery profiles across the threshold range, with ARI values approximately equal to 0.943 and 0.914, respectively. This indicates that, in the simulated setting, the recovered partitions are not artifacts of the specific 95\% truncation rule. Rather, the main latent structure is already stable over the broader 90\% to 99\% interval.

\subsection{Interpretative case study: identifying temporal risk dynamics}

To complement the continuum analysis, we consider a representative simulated scenario in which the true latent structure consists of three temporal risk regimes. The objective of this case study is not only to compare numerical recovery, but also to illustrate how the proposed FPCA-based log-hazard pipeline represents and interprets the temporal structure of the simulated risk trajectories. As described in the simulation design, the data are generated at the subject level and then aggregated into curve-level cohorts, yielding smoothed log-hazard trajectories on which FPCA and clustering are performed.

Standard FPCA was applied to the B-spline smoothed log-hazard curves, and clustering was performed directly on the resulting unstandardized FPCA scores. In this simulation, the FPCA truncation rule selected \(M=8\) retained components. Since the true partition is known by construction, the Adjusted Rand Index is used as the primary external recovery measure, while the average Silhouette width provides a supporting internal measure of geometric cohesion and separation. Stability is reported for the FPCA-based clustering methods as an additional diagnostic of partition robustness.

\begin{table*}[!ht]
\centering
\caption{Clustering performance comparison on the simulated subject-level survival data. ARI is computed with respect to the true latent partition. Silhouette is computed in the corresponding clustering space and should therefore be interpreted as an internal diagnostic within each representation.}
\label{tab:sim_metrics}
\resizebox{\textwidth}{!}{
\begin{tabular}{lcccc}
\hline
\textbf{Clustering Method} & \textbf{Selected \(K\)} & \textbf{ARI} & \textbf{Silhouette} & \textbf{Stability} \\
\hline
\textbf{K-means on FPCA scores} & \textbf{3} & \textbf{1.000} & 0.279 & 0.985 \\
K-medoids on FPCA scores & 3 & 0.943 & 0.275 & 0.959 \\
Ward clustering on FPCA scores & 3 & 0.914 & 0.268 & 0.916 \\
K-means on standardized B-spline coefficients & 3 & 0.943 & 0.091 & -- \\
Survival probability curve benchmark & 3 & 0.351 & 0.362 & -- \\
\hline
\end{tabular}
}
\end{table*}

The results in Table~\ref{tab:sim_metrics} show that clustering in the FPCA score space accurately recovers the latent temporal regimes. K-means on unstandardized FPCA scores achieves perfect recovery of the true partition, with ARI equal to 1.000 and high stability equal to 0.985. K-medoids and Ward clustering also recover the latent structure with high ARI values, equal to 0.943 and 0.914, respectively. The coefficient-based benchmark performs well in terms of ARI, but its low Silhouette value indicates weak geometric cohesion in the standardized coefficient space. By contrast, the survival probability curve benchmark selects the correct number of groups but yields a much lower ARI, equal to 0.351, indicating that the cumulative survival representation does not recover the time-localized log-hazard regimes generated by the simulation.

This discrepancy is important. The survival probability benchmark can produce internally cohesive groups in the survival space, as reflected by its Silhouette value, but these groups do not correspond well to the latent risk regimes defined on the log-hazard scale. This confirms that internal cohesion in cumulative survival curves does not necessarily imply recovery of the underlying instantaneous risk dynamics.

\begin{figure*}[!ht]
    \centering
    \begin{minipage}[t]{0.49\textwidth}
        \centering
        \includegraphics[width=\textwidth]{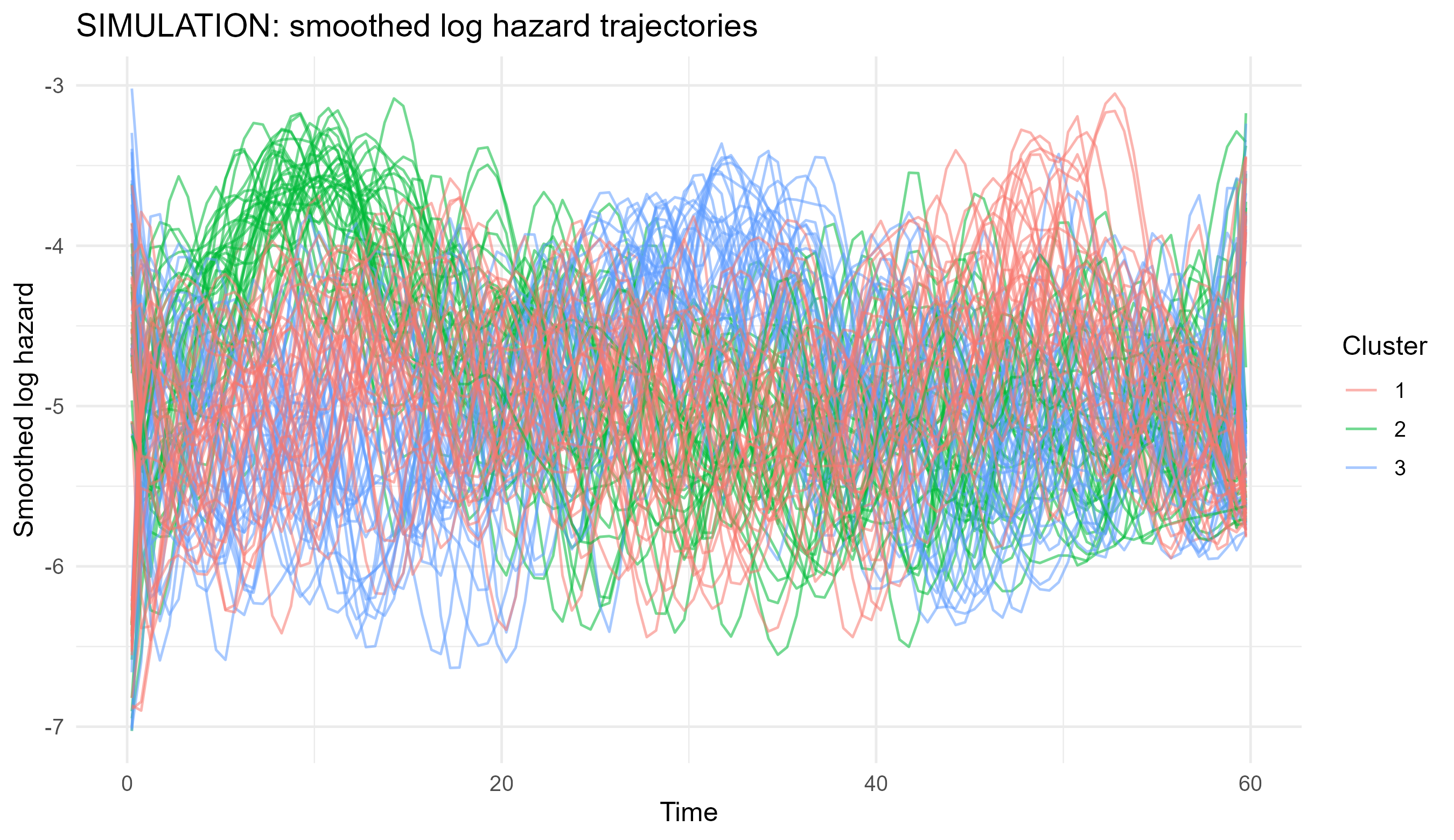}
        \caption{Simulated smoothed log-hazard trajectories clustered in the FPCA score space. The three colors identify the clusters recovered by the proposed functional pipeline.}
        \label{fig:sim_loghazard_clusters}
    \end{minipage}
    \hfill
    \begin{minipage}[t]{0.49\textwidth}
        \centering
        \includegraphics[width=\textwidth]{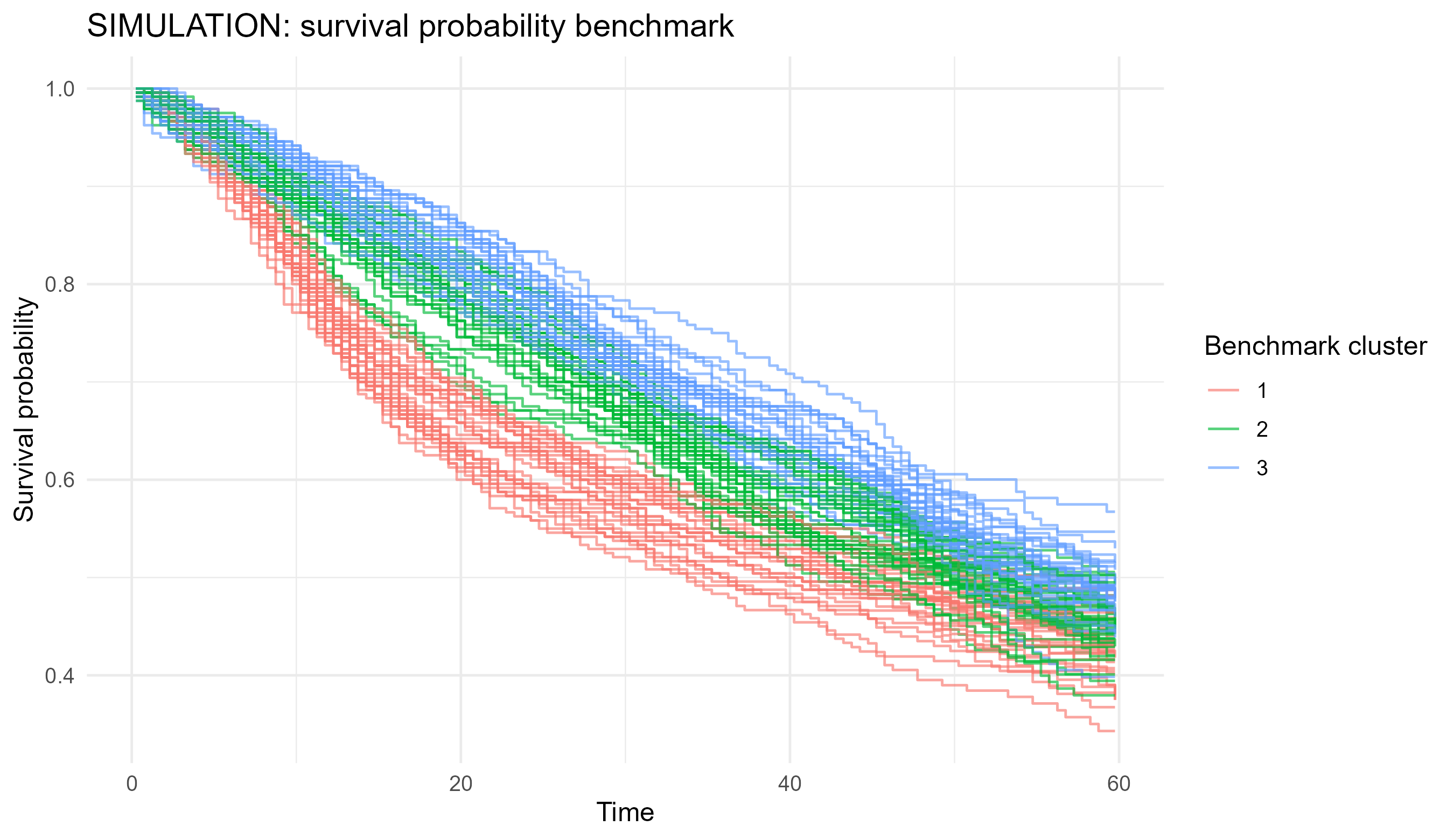}
        \caption{Simulation benchmark based on cumulative survival probability curves. Although three benchmark groups are identified, the cumulative survival representation does not recover the latent log-hazard regimes with the same accuracy.}
        \label{fig:sim_survival_benchmark}
    \end{minipage}
\end{figure*}

Figures~\ref{fig:sim_loghazard_clusters} and~\ref{fig:sim_survival_benchmark} provide a visual explanation of the recovery results. In the log-hazard domain, the three recovered groups correspond to distinct temporal risk patterns, although individual trajectories remain noisy and partially overlapping. In the cumulative survival domain, the trajectories are more strongly compressed by integration over time, and the resulting benchmark groups primarily reflect broad survival decline rather than the timing and localization of risk variation.

\begin{figure*}[!ht]
    \centering
    \includegraphics[width=0.85\textwidth]{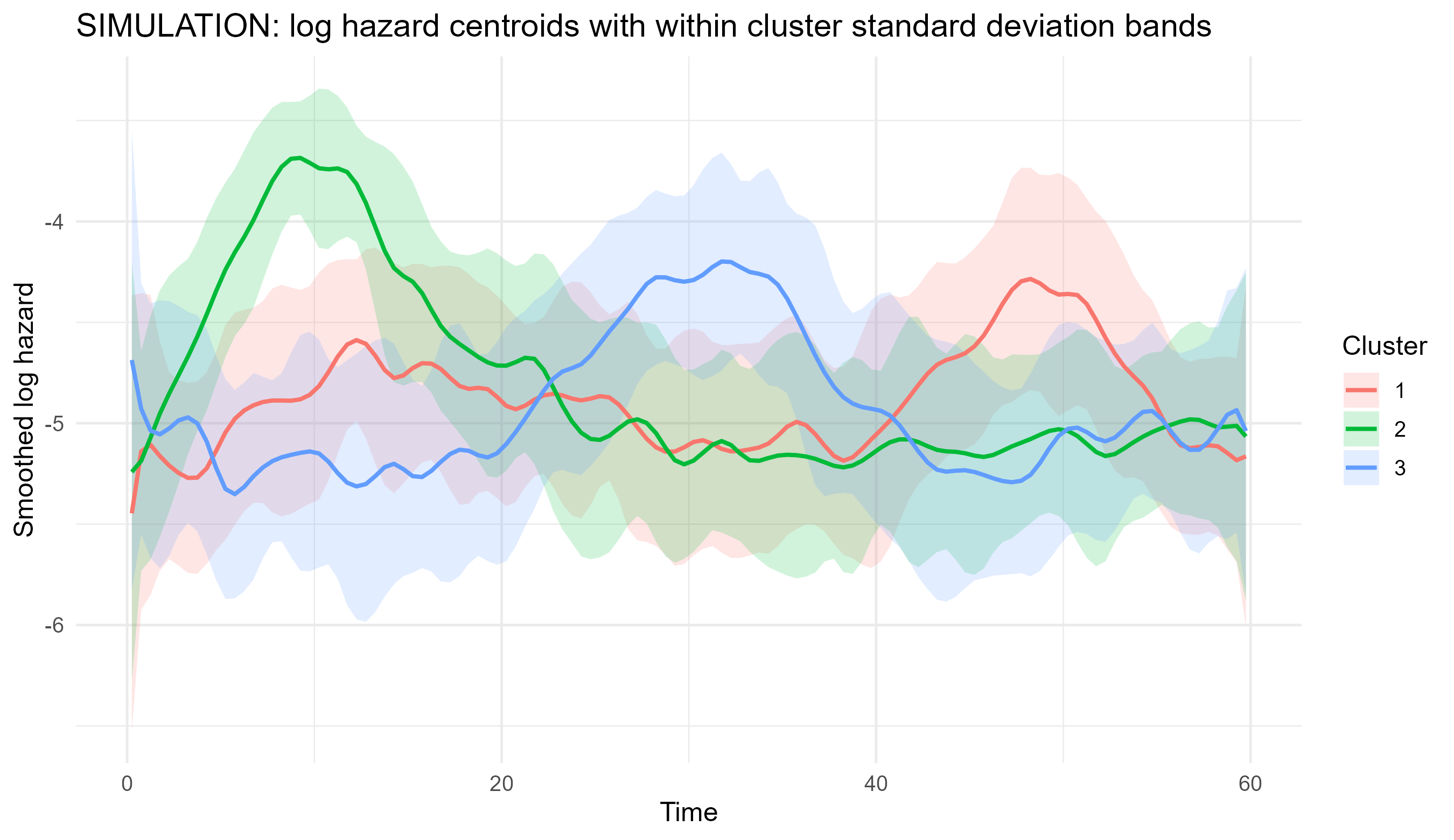}
    \caption{Estimated log-hazard centroids for the three simulated functional clusters. The shaded bands represent within-cluster standard deviation and describe the functional dispersion of the risk dynamics along the time domain, not confidence intervals for the mean.}
    \label{fig:sim_centroids}
\end{figure*}

To interpret the recovered partition in the time domain, Figure~\ref{fig:sim_centroids} reports the cluster centroids with within-cluster standard deviation bands. The three centroid curves summarize distinct temporal profiles. One group is characterized by an early elevation in the log-hazard, another by an intermediate increase, and the third by a later risk elevation. The variability bands indicate that the groups are not merely separated by a constant vertical shift, but also differ in the timing and persistence of risk fluctuations. This supports the use of the log-hazard representation as a functional object, since the relevant differences are distributed over the temporal domain.

\begin{figure*}[!ht]
    \centering
    \begin{minipage}[t]{0.49\textwidth}
        \centering
        \includegraphics[width=\textwidth]{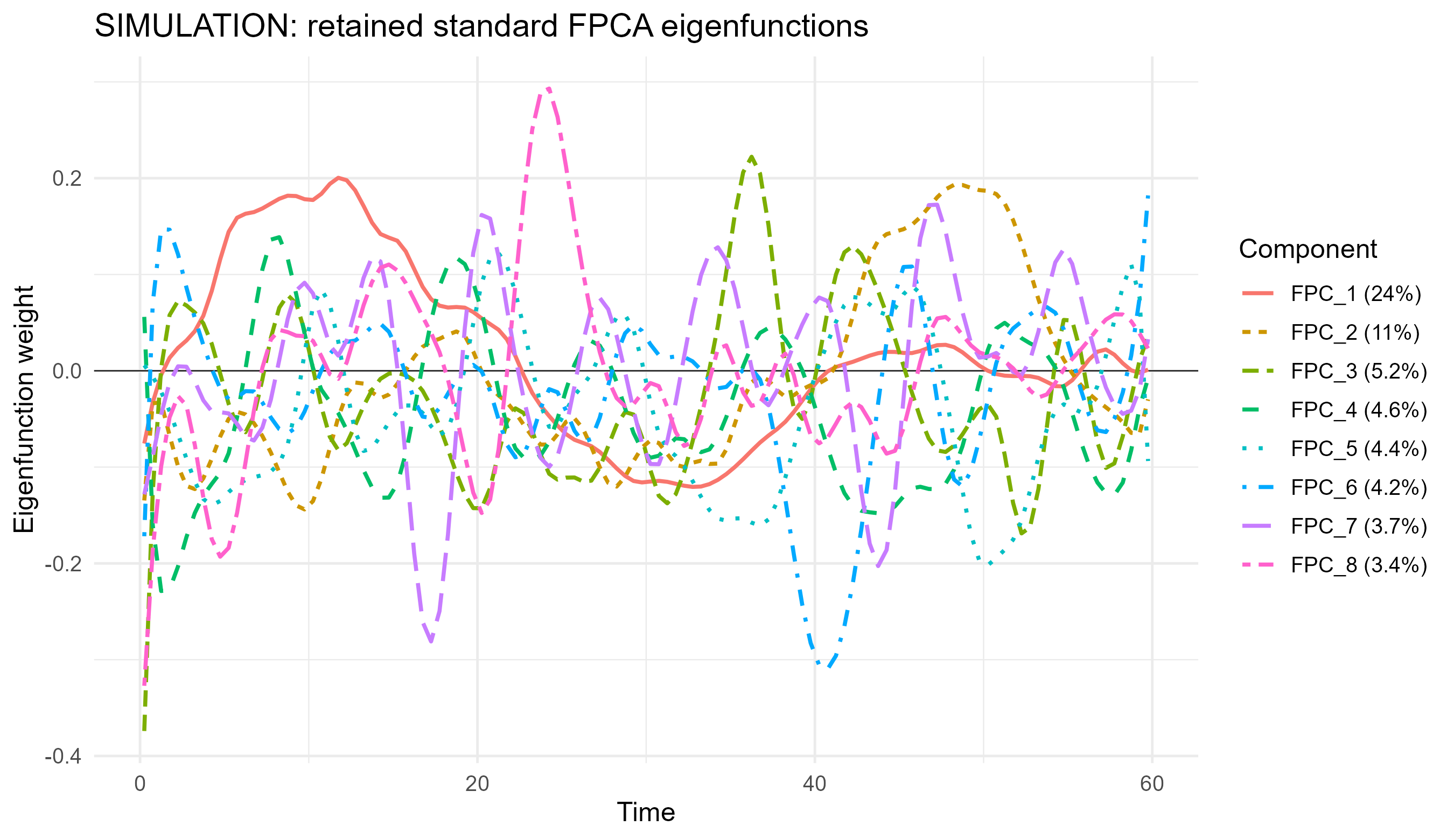}
        \caption{Retained standard FPCA eigenfunctions extracted from the smoothed log-hazard trajectories in the simulation study. The legend reports the proportion of functional variance associated with each component.}
        \label{fig:sim_eigenfunctions}
    \end{minipage}
    \hfill
    \begin{minipage}[t]{0.49\textwidth}
        \centering
        \includegraphics[width=\textwidth]{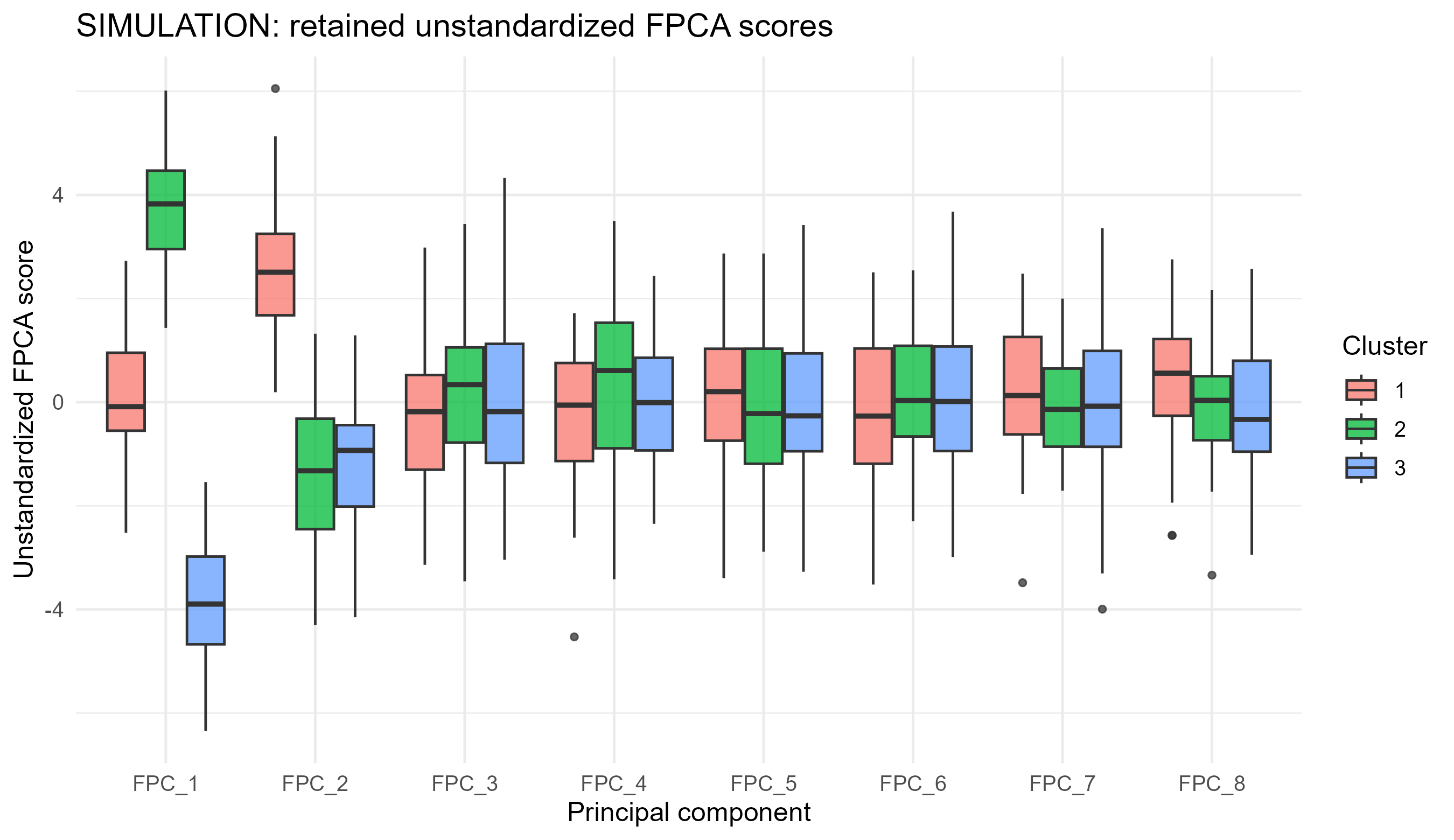}
        \caption{Distribution of the retained unstandardized FPCA scores by recovered cluster. The score distributions show which components contribute most strongly to the separation and within-cluster variability of the simulated log-hazard trajectories.}
        \label{fig:sim_scores}
    \end{minipage}
\end{figure*}

The post-hoc FPCA interpretation in Figures~\ref{fig:sim_eigenfunctions} and~\ref{fig:sim_scores} links the geometric clustering result back to the functional domain. The first components capture broad differences in the level and timing of log-hazard variation, while later components describe more localized temporal fluctuations. The score distributions show that the main between-cluster separation is concentrated in the first retained components, especially FPC 1 and FPC 2, whereas later components mainly contribute to residual within-cluster heterogeneity. This confirms that the clustering is driven by interpretable modes of temporal risk variation rather than by arbitrary pointwise noise.

This simulation case study demonstrates that the proposed FPCA-based log-hazard clustering framework can recover time-localized risk regimes from right-censored survival data. Rather than identifying definitive clinical phenotypes, the simulation shows that the method can isolate distinct temporal patterns of instantaneous risk that may be obscured when the same data are represented only through cumulative survival probabilities.

\subsection{Stress-test: robustness to outlier trajectories}

A critical issue in clustering survival trajectories, even in controlled simulation settings, is the presence of heterogeneous or extreme curves, which can disproportionately attract cluster centroids and distort the resulting partition. To evaluate the robustness of our framework against such anomalies, we conducted an additional stress-test by incrementally injecting functionally anomalous curves into the simulated cohorts. The recovery metric was computed only on the original baseline curves, so that the analysis evaluates whether the injected outliers compromise the recovery of the original latent partition. For each outlier percentage, the number of retained FPCs was selected by the same FPCA truncation rule used in the main pipeline.

\begin{table*}[!ht]
\centering
\caption{Outlier stress-test in the simulation study. ARI is computed on the original baseline curves only, while Silhouette is computed on all curves after outlier injection. The column \(M\) reports the number of retained FPCs selected by the pipeline.}
\label{tab:stress_test_outliers}
\resizebox{\textwidth}{!}{
\begin{tabular}{c l c c c c}
\hline
\textbf{Outliers} & \textbf{Method} & \textbf{Selected \(K\)} & \textbf{ARI} & \textbf{Silhouette} & \textbf{Selected \(M\)} \\
\hline
0\%  & K-means on FPCA scores & 2 & 0.553 & 0.282 & 8 \\
0\%  & K-medoids on FPCA scores & 3 & 0.862 & 0.266 & 8 \\
0\%  & Ward clustering on FPCA scores & 2 & 0.519 & 0.275 & 8 \\
0\%  & K-means on standardized B-spline coefficients & 2 & 0.540 & 0.091 & 8 \\
\hline
10\% & K-means on FPCA scores & 2 & 0.540 & 0.270 & 8 \\
10\% & K-medoids on FPCA scores & 2 & 0.443 & 0.250 & 8 \\
10\% & Ward clustering on FPCA scores & 2 & 0.553 & 0.264 & 8 \\
10\% & K-means on standardized B-spline coefficients & 2 & 0.476 & 0.101 & 8 \\
\hline
20\% & K-means on FPCA scores & 2 & 0.567 & 0.254 & 8 \\
20\% & K-medoids on FPCA scores & 2 & 0.567 & 0.254 & 8 \\
20\% & Ward clustering on FPCA scores & 2 & 0.567 & 0.254 & 8 \\
20\% & K-means on standardized B-spline coefficients & 2 & 0.567 & 0.100 & 8 \\
\hline
30\% & K-means on FPCA scores & 4 & 0.943 & 0.245 & 8 \\
30\% & K-medoids on FPCA scores & 3 & 0.943 & 0.229 & 8 \\
30\% & Ward clustering on FPCA scores & 6 & 0.804 & 0.225 & 8 \\
30\% & K-means on standardized B-spline coefficients & 2 & 0.553 & 0.086 & 8 \\
\hline
\end{tabular}
}
\end{table*}

Table~\ref{tab:stress_test_outliers} shows that the effect of outlier contamination is not monotone and depends on the interaction between the injected anomalous trajectories, the selected number of clusters, and the clustering method. At 0\% outliers, K-medoids provides the best recovery among the evaluated methods, with ARI equal to 0.862, whereas K-means and Ward clustering select two clusters and show lower agreement with the original three-group structure. When 10\% and 20\% outliers are injected, the methods tend to select a more parsimonious two-cluster structure, leading to moderate ARI values. At 30\% outliers, K-means and K-medoids recover the original baseline partition more accurately, with ARI equal to 0.943, although this occurs together with a change in the selected number of clusters.

These results suggest that the outlier stress-test should be interpreted as a robustness analysis of the complete FPCA-clustering pipeline, rather than as an isolated comparison of clustering algorithms at fixed \(K\). Since \(K\) is reselected after outlier injection, changes in ARI reflect the joint effect of anomalous curves, FPCA score geometry, and cluster-number selection. In this sense, the stress-test indicates that FPCA-based clustering can remain informative under contamination, but it also shows that the selected partition may be sensitive to the way outliers alter the global geometry of the score space.

The coefficient-based benchmark remains comparatively weak in terms of internal cohesion across all contamination levels, with Silhouette values close to 0.10 or below. This supports the use of the FPCA score representation rather than direct clustering of standardized B-spline coefficients. However, the results also caution against presenting any single clustering method as uniformly superior under all forms of contamination. The main methodological conclusion is that the FPCA representation provides a structured and interpretable space in which different clustering algorithms can be compared and stress-tested under controlled perturbations.

\section{Real-World case studies}\label{sec.real_data}

To investigate the behavior of the proposed approach in practical scenarios, we applied the functional log-hazard clustering framework to two real clinical datasets. Since no ground-truth partition is available in these observational applications, the comparison focuses on internal cohesion, stability diagnostics, sensitivity to the FPCA truncation threshold, and functional interpretability rather than on external recovery measures.

The proposed method is compared with two non-functional reference representations. The first is a coefficient-based benchmark, obtained by clustering standardized B-spline coefficient vectors. The second is a cumulative-survival benchmark, obtained by applying clustering methods to discretized survival probability trajectories. This comparison is designed to assess whether clustering smoothed log-hazard trajectories in the FPCA score space provides more interpretable and internally coherent partitions than clustering either spline coefficients directly or cumulative survival probability curves.

\subsection{The German Breast Cancer Study (GBCS)}

The first real-world application utilizes the German Breast Cancer Study (GBCS) dataset, available as \textit{gbcsCS} in the R \textit{condSURV} package. In this dataset, patient groups are formed based on the number of affected lymph nodes, which ranges from $1$ to $13+$. This serves as an initial form of clinical stratification, providing a biologically meaningful way to categorize patients, as also noted by \cite{RJ-2021-032}. 

For the comparative analysis, temporal trajectories for each of the $14$ clinical cohorts were constructed using recurrence times and right-censoring indicators. The proposed FDA pipeline extracts empirical log-hazard trajectories, smooths them through the GCV-constrained B-spline rule described in Section~\ref{sec.2}, applies standard FPCA to the smoothed log-hazard functions, and performs clustering on the resulting unstandardized FPCA scores. Since an absolute ground-truth partition is inherently absent in observational clinical data, the resulting clusters are evaluated using internal cohesion, stability diagnostics, sensitivity analysis, and functional interpretability rather than external recovery metrics such as the Adjusted Rand Index.

For GBCS, the GCV-constrained smoothing rule selected a B-spline configuration with \(n_\beta=60\) and \(\lambda=0.1\). The selected smoother remained close to the GCV optimum, with a relative GCV increase of 2.1\%, while allowing greater local flexibility in the reconstructed log-hazard trajectories. The FPCA truncation rule retained \(M=7\) components, explaining 96.4\% of the total functional variance, thereby satisfying the 95\% cumulative explained-variance criterion. The first FPC alone explained 79.2\% of the functional variance, while the following retained components captured smaller but non-negligible temporal shape variations.

\begin{table*}[!ht]
\centering
\caption{Clustering comparison on the GBCS dataset. The proposed method clusters the unstandardized FPCA scores of the smoothed log-hazard trajectories. The coefficient baseline clusters standardized B-spline coefficient vectors, while the survival benchmark clusters discretized cumulative survival probability trajectories.}
\label{tab:results_gbcs_integrated}
\begin{tabular}{lccc}
\hline
\textbf{Representation and method} & \textbf{Selected \(K\)} & \textbf{Selected \(M\)} & \textbf{Silhouette} \\
\hline
\textbf{K-means on unstandardized FPCA scores} & \textbf{2} & \textbf{7} & \textbf{0.484} \\
K-means on standardized B-spline coefficients & 2 & \(\mathrm{NA}\) & 0.413 \\
K-means on discretized survival probability curves & 3 & \(\mathrm{NA}\) & 0.521 \\
\hline
\end{tabular}
\end{table*}

Table~\ref{tab:results_gbcs_integrated} shows that the selected functional solution is a two-cluster partition obtained by K-means in the unstandardized FPCA score space, with average Silhouette width equal to 0.484. The coefficient-based baseline also selects \(K=2\), but with a lower Silhouette value of 0.413, indicating weaker cohesion in the standardized coefficient representation. The cumulative-survival benchmark selects \(K=3\) and attains a Silhouette value of 0.521. This higher value should not be interpreted as external superiority, because no ground-truth partition is available in the GBCS application and the Silhouette is computed in a different representation space. Rather, it indicates that the survival probability curves admit an internally cohesive partition in the cumulative-survival domain, whereas the proposed method targets a different object, namely the temporal dynamics of the smoothed log-hazard.

\begin{figure*}[!t]
    \centering
    \begin{minipage}[t]{0.49\textwidth}
        \centering
        \includegraphics[width=\textwidth]{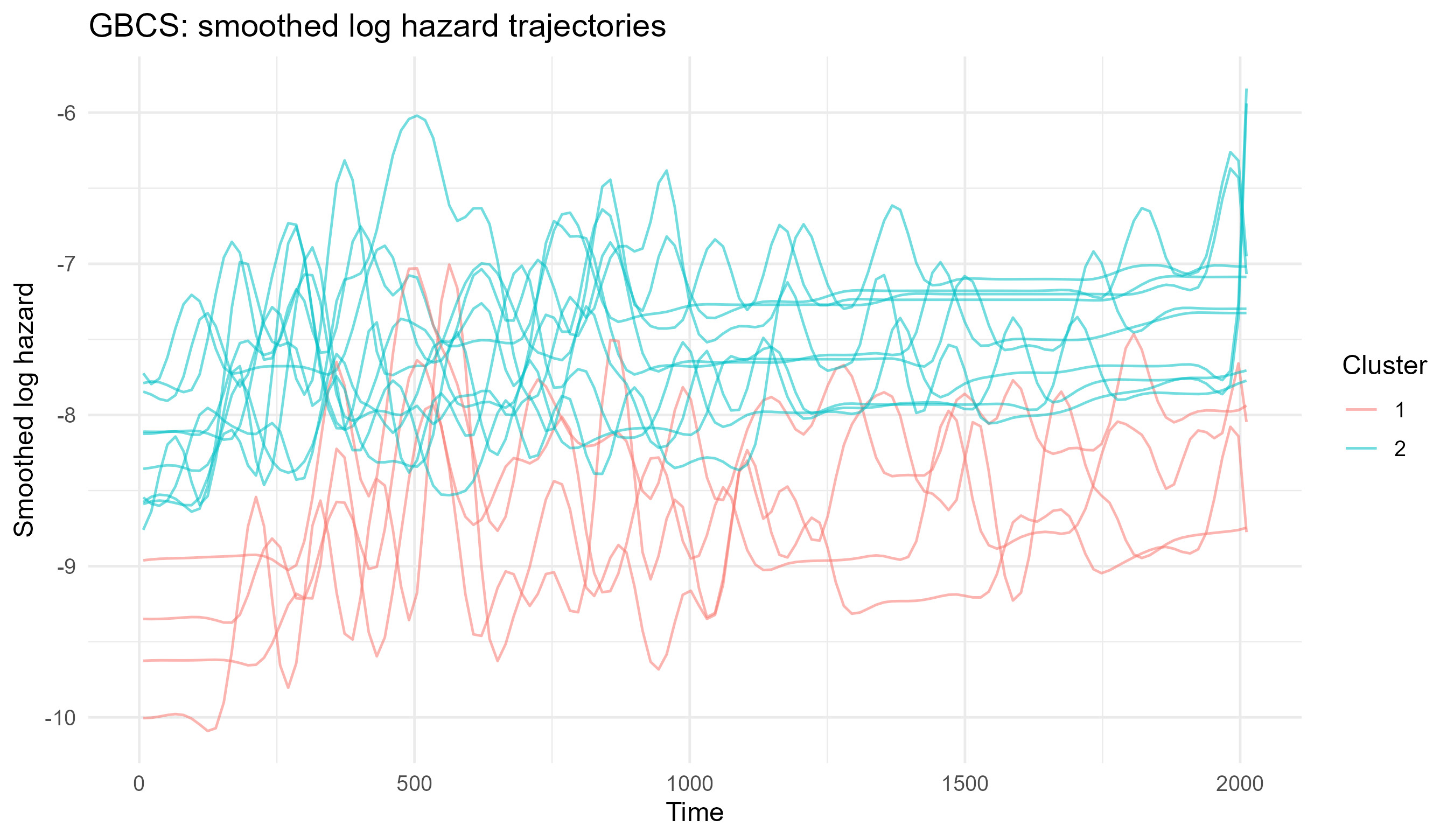}
        
        \smallskip
        \textbf{(a)} Smoothed log-hazard trajectories clustered in the FPCA score space.
    \end{minipage}
    \hfill
    \begin{minipage}[t]{0.49\textwidth}
        \centering
        \includegraphics[width=\textwidth]{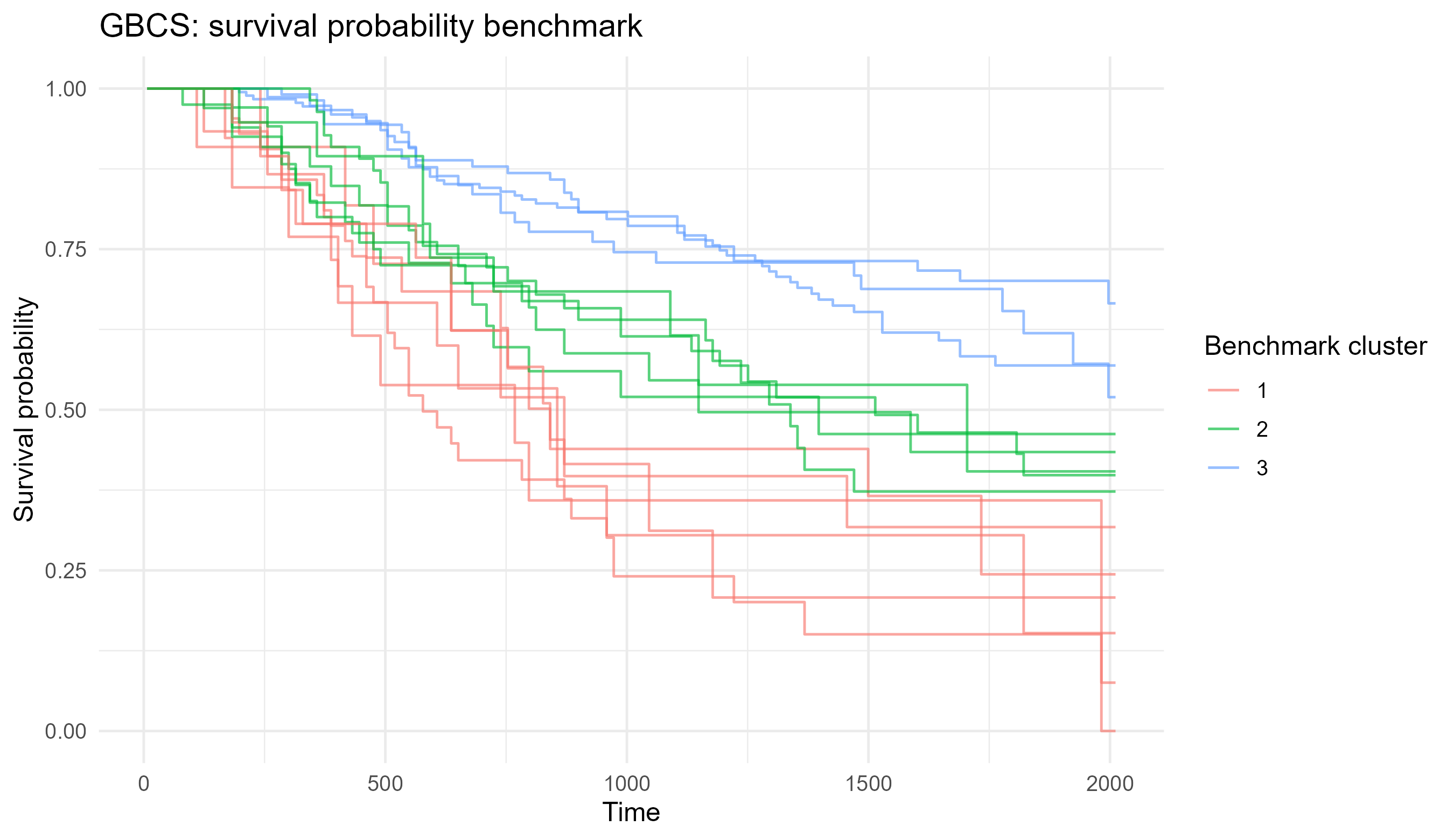}
        
        \smallskip
        \textbf{(b)} Benchmark clustering on discretized survival probability trajectories.
    \end{minipage}
    \caption{GBCS clustering comparison between the proposed functional log-hazard representation and the cumulative-survival benchmark. Panel (a) reports the smoothed log-hazard trajectories clustered in the unstandardized FPCA score space. Panel (b) reports the benchmark partition obtained from discretized survival probability curves. The two representations answer related but distinct questions: the proposed method groups cohorts according to the temporal dynamics of instantaneous recurrence risk, whereas the survival benchmark groups cohorts according to cumulative survival decline.}
    \label{fig:gbcs_comparison}
\end{figure*}

Figure~\ref{fig:gbcs_comparison} illustrates the difference between the two representations. In the log-hazard domain, the selected functional partition separates the \(14\) node-based cohorts into two groups with visibly different levels and temporal profiles of instantaneous recurrence risk. In the cumulative-survival domain, the benchmark produces three groups that mainly reflect the progressive separation of survival probability curves over follow-up time. These two partitions answer related but distinct questions: the survival benchmark groups cohorts according to cumulative event history, whereas the functional log-hazard approach groups cohorts according to the temporal evolution of instantaneous risk.

The functional cluster composition is clinically interpretable but should be treated as exploratory. Cluster 1 contains the cohorts with 1, 2, 3, and 4 positive lymph nodes. Cluster 2 contains the cohorts with 5, 6, 7, 8, 9, 10, 11, 12, 13, and \(14+\) positive lymph nodes. Therefore, in this application the functional partition identifies a threshold-like separation between lower nodal involvement and higher nodal involvement. This pattern is consistent with the expected prognostic role of lymph node burden, but it should not be interpreted as an externally validated clinical cutoff. Rather, it indicates that the smoothed log-hazard trajectories of the lower-node cohorts are closer to each other in the FPCA score space than to those of the higher-node cohorts.

\begin{figure*}[!t]
    \centering
    \includegraphics[width=0.85\textwidth]{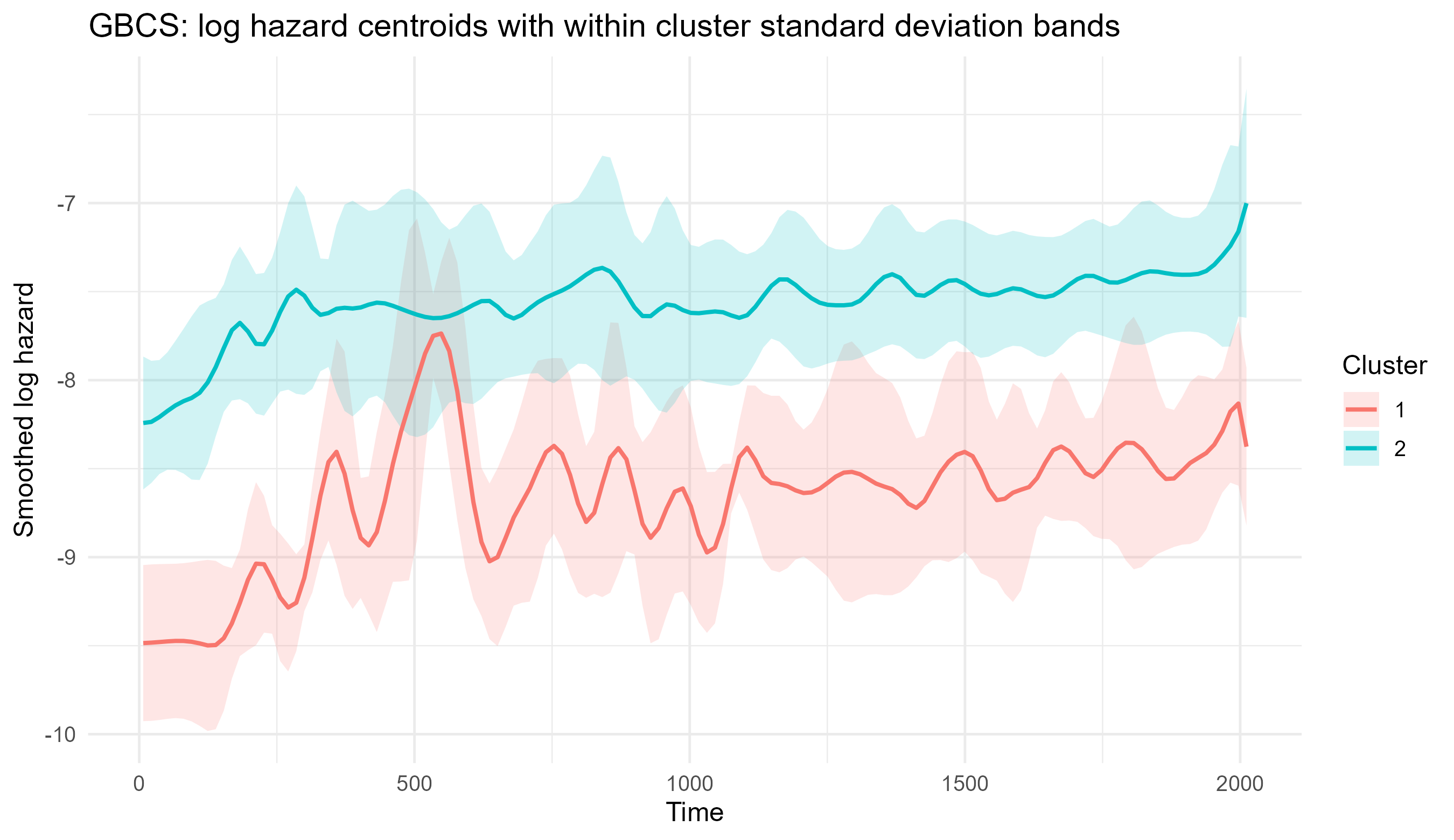}
    \caption{GBCS log-hazard centroids with within-cluster standard deviation bands. The shaded regions describe functional dispersion around each centroid and should not be interpreted as confidence intervals for the mean.}
    \label{fig:gbcs_centroids}
\end{figure*}

The centroid representation in Figure~\ref{fig:gbcs_centroids} clarifies the temporal nature of the two groups. Cluster 1 is characterized by lower log-hazard levels across most of the follow-up period, whereas Cluster 2 exhibits a systematically higher log-hazard trajectory. The separation is therefore driven primarily by a persistent level difference in instantaneous recurrence risk. The standard deviation bands indicate non-negligible within-cluster heterogeneity, especially in regions where the empirical hazard reconstruction is more variable, but the two centroid curves remain clearly separated over most of the time domain.

\begin{figure*}[!t]
    \centering
    \begin{minipage}[t]{0.49\textwidth}
        \centering
        \includegraphics[width=\textwidth]{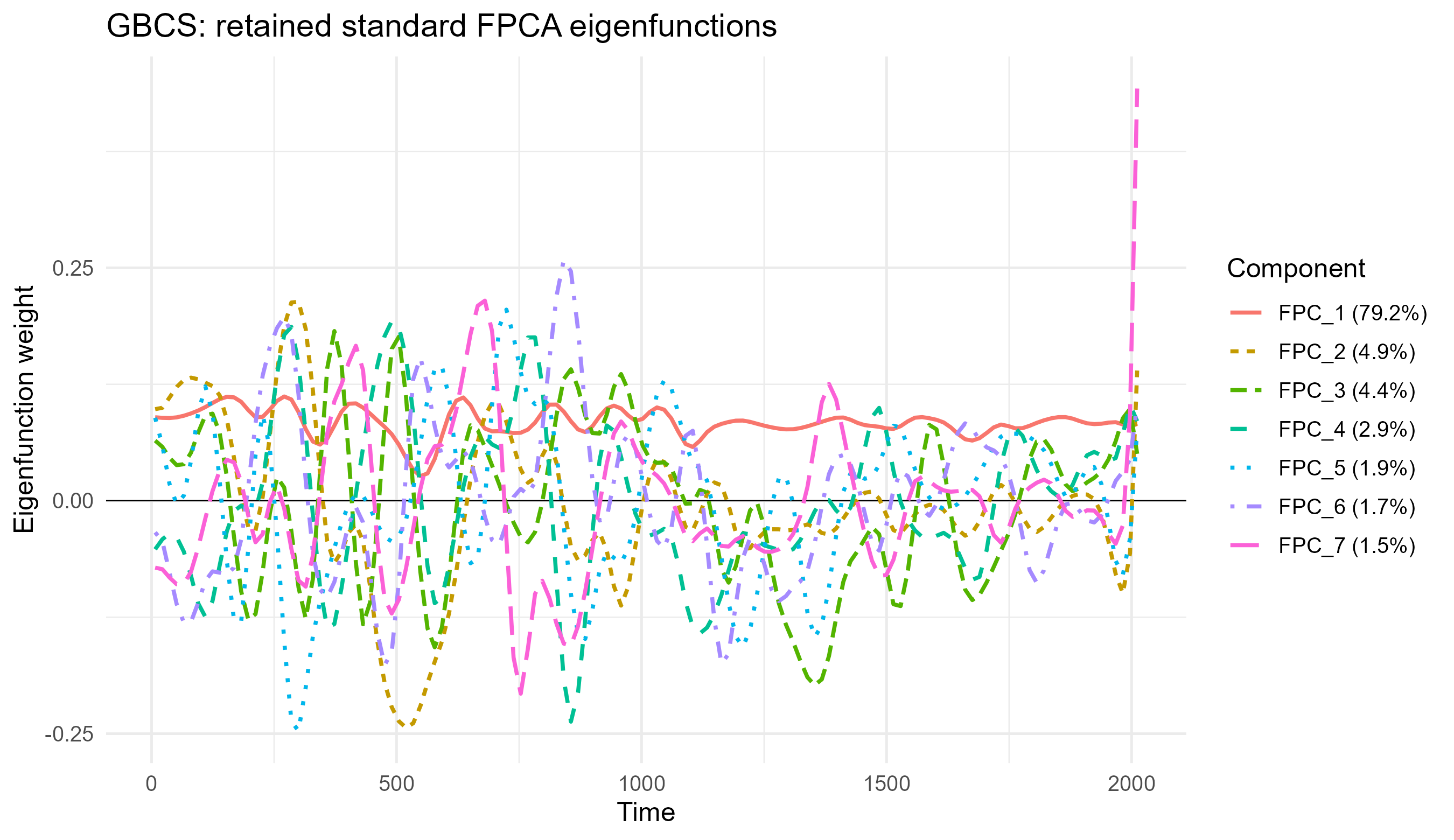}
        
        \smallskip
        \textbf{(a)} Retained standard FPCA eigenfunctions.
    \end{minipage}
    \hfill
    \begin{minipage}[t]{0.49\textwidth}
        \centering
        \includegraphics[width=\textwidth]{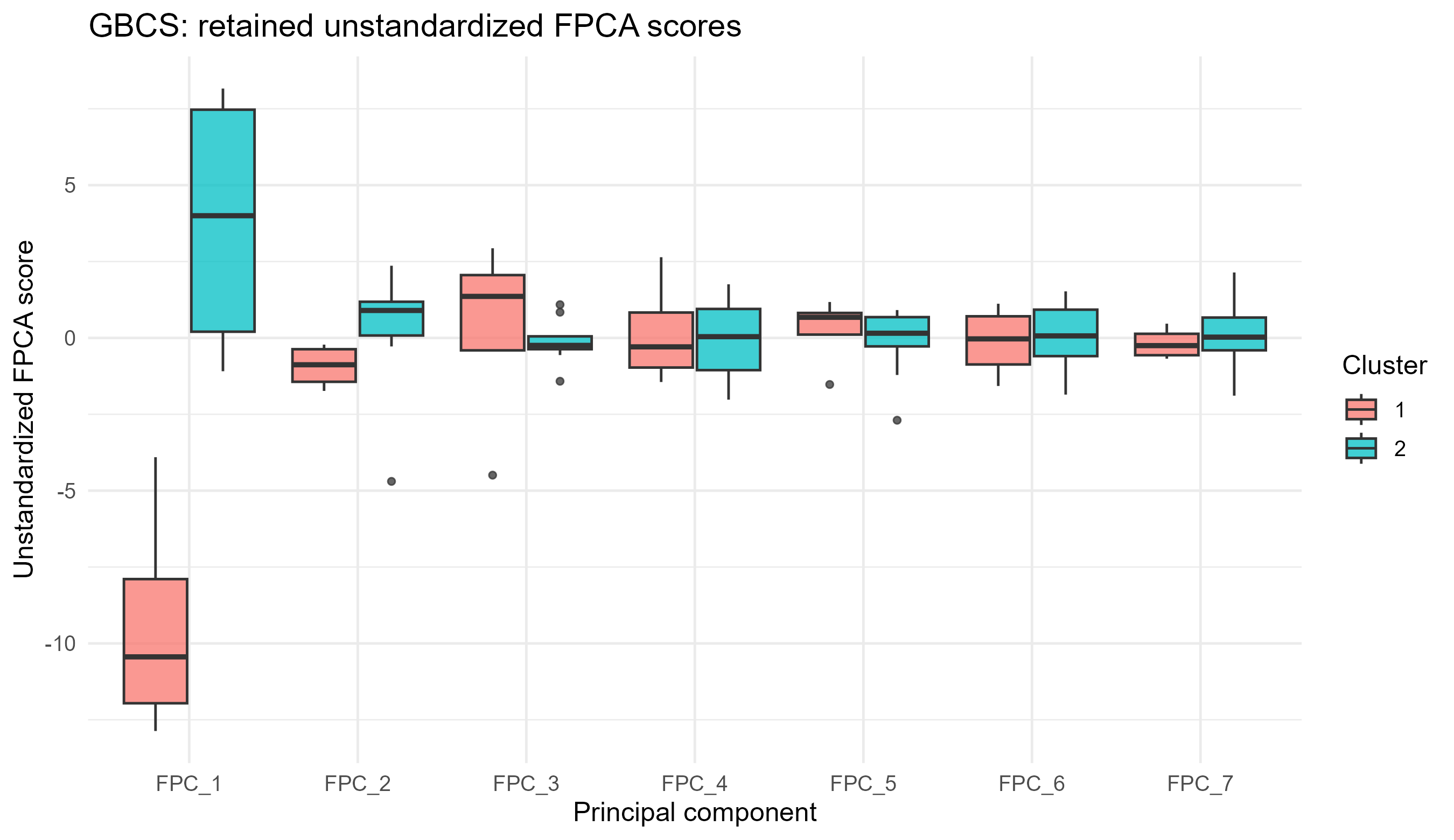}
        
        \smallskip
        \textbf{(b)} Retained unstandardized FPCA scores by cluster.
    \end{minipage}
    \caption{Post-hoc FPCA interpretation for the GBCS application. Panel (a) reports the retained standard FPCA eigenfunctions extracted from the smoothed log-hazard trajectories, with the proportion of functional variance explained by each component. Panel (b) reports the distribution of the retained unstandardized FPCA scores by cluster, showing which components contribute most strongly to between-cluster separation and within-cluster heterogeneity.}
    \label{fig:gbcs_fpca_explainability}
\end{figure*}

Figure~\ref{fig:gbcs_fpca_explainability} provides a post-hoc interpretation of the partition in the FPCA space. FPC 1 explains 79.2\% of the total functional variance and is the dominant direction of separation between the two clusters. The corresponding score distributions show a clear separation along FPC 1, with Cluster 1 located at substantially lower FPC 1 scores and Cluster 2 at higher scores. This confirms that the main partition is driven by a broad level difference in the smoothed log-hazard trajectories. The remaining retained components explain smaller proportions of variance, namely 4.9\%, 4.4\%, 2.9\%, 1.9\%, 1.7\%, and 1.5\%, and mainly capture local temporal deviations and within-cluster variability rather than the primary between-cluster separation.

Finally, bootstrap stability was used as an internal diagnostic of partition robustness conditional on the estimated functional trajectories. The mean Jaccard indices were 0.599 for Cluster 1 and 0.601 for Cluster 2, with an overall mean Jaccard value of 0.600. These values indicate moderate, rather than excellent, stability. Therefore, the GBCS partition should be interpreted as an internally coherent and functionally interpretable exploratory structure, but not as definitive evidence of a uniquely stable clinical classification. A full assessment of clinical validity would require independent data or a resampling scheme that propagates the uncertainty of the original survival and hazard estimation process.

\subsection{The Primary Biliary Cirrhosis (PBC) Dataset}

To further assess the behavior of the proposed functional framework in a second clinical application, we analyzed the Primary Biliary Cirrhosis (PBC) dataset from the Mayo Clinic \citep{therneau2000modeling}. The main aim of this application is exploratory: the aim is to evaluate whether clustering smoothed log-hazard trajectories can identify interpretable temporal risk profiles in a setting different from the GBCS recurrence data. Since no external ground-truth partition is available, the analysis is based on internal cohesion, stability diagnostics, sensitivity to the FPCA truncation rule, and functional interpretability.

Patients were stratified according to two clinically relevant axes: histological stage of the disease and serum bilirubin concentration. Histological stage was considered from Stage 1 to Stage 4, while bilirubin concentration was discretized into quartiles. Composite clinical profiles with insufficient sample size were excluded in order to obtain more stable empirical hazard estimates. This yielded \(N=12\) aggregate temporal trajectories. The composite endpoint was defined as death or liver transplantation.

For PBC, the GCV-constrained smoothing rule selected a B-spline configuration with \(n_\beta=40\) and \(\lambda=0.051507\). The selected smoother remained close to the GCV optimum, with a relative GCV increase of 2.1\%, while preserving local flexibility in the reconstructed log-hazard trajectories. The FPCA truncation rule retained \(M=2\) components, explaining 95.5\% of the total functional variance and therefore satisfying the 95\% cumulative explained-variance criterion. The first FPC alone explained 93.4\% of the functional variance, while the second FPC explained an additional 2.1\%, mainly capturing residual temporal shape variation.
\begin{table*}[!ht]
\centering
\caption{Comparison of clustering approaches on the PBC dataset. The proposed FDA framework clusters unstandardized FPCA scores obtained from smoothed log-hazard trajectories. In contrast, the coefficient-based baseline relies on standardized B-spline coefficient vectors, while the survival-based benchmark clusters discretized cumulative survival probability trajectories.}

\label{tab:results_pbc_integrated}
\begin{tabular}{lccc}
\hline
\textbf{Representation and method} & \textbf{Selected \(K\)} & \textbf{Selected \(M\)} & \textbf{Silhouette} \\
\hline
\textbf{K-means on unstandardized FPCA scores} & \textbf{2} & \textbf{2} & \textbf{0.526} \\
K-means on standardized B-spline coefficients & 2 & \(\mathrm{NA}\) & 0.426 \\
K-means on discretized survival probability curves & 2 & \(\mathrm{NA}\) & 0.543 \\
\hline
\end{tabular}
\end{table*}

Table~\ref{tab:results_pbc_integrated} shows that the proposed functional solution selects a two-cluster partition in the unstandardized FPCA score space, with an average Silhouette width equal to 0.526. The coefficient-based baseline also selects \(K=2\), but with a lower Silhouette value of 0.426, indicating weaker internal cohesion in the standardized coefficient space. The cumulative-survival benchmark also selects \(K=2\) and attains a Silhouette value of 0.543. As in the GBCS application, this value should not be interpreted as external superiority, since the Silhouette is computed in a different representation space and no true partition is available. Rather, it indicates that the cumulative survival trajectories admit a cohesive binary partition, whereas the proposed method targets the temporal dynamics of instantaneous risk on the smoothed log-hazard scale.

\begin{figure*}[!t]
    \centering
    \begin{minipage}[t]{0.49\textwidth}
        \centering
        \includegraphics[width=\textwidth]{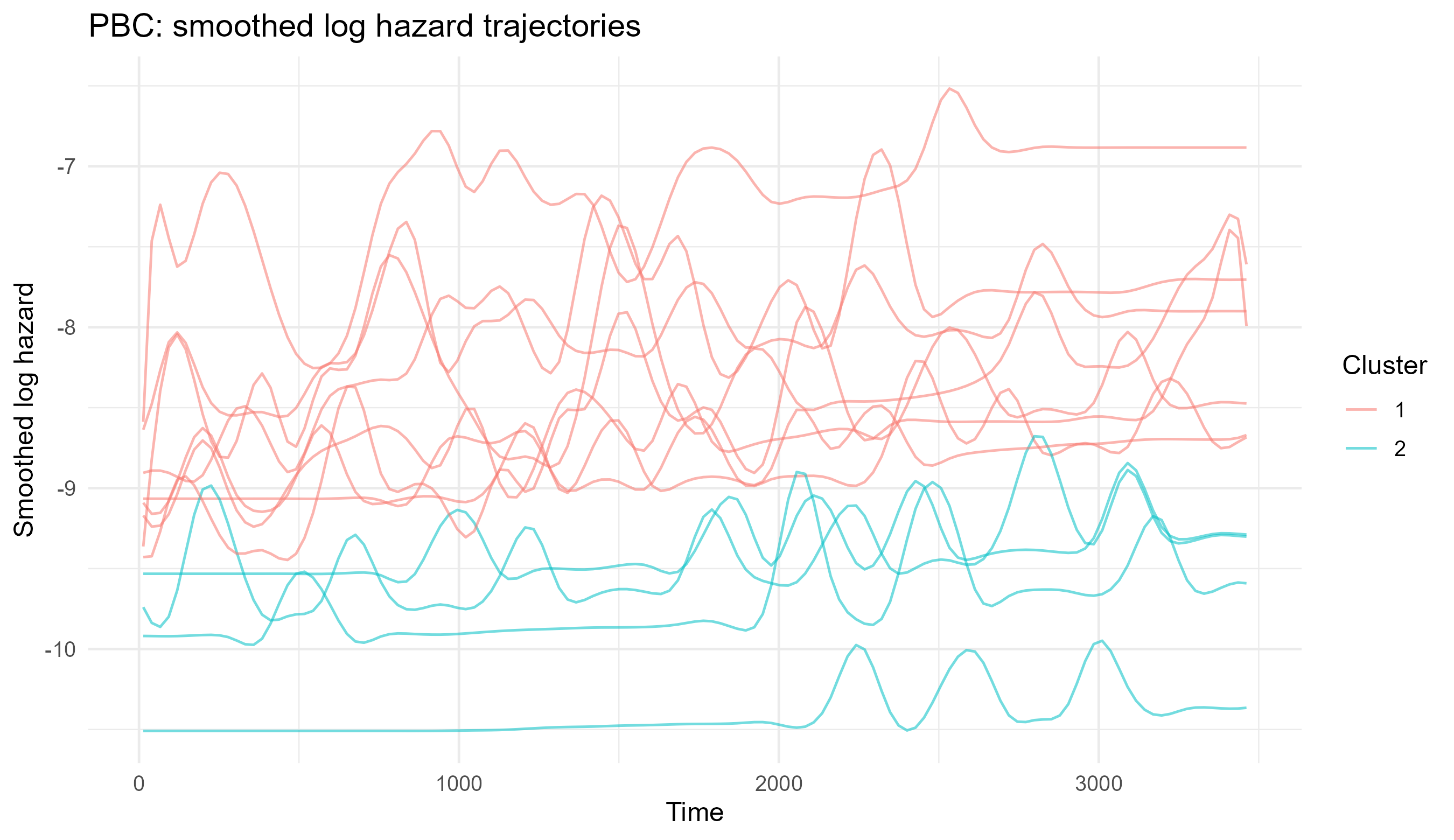}
        
        \smallskip
        \textbf{(a)} Smoothed log-hazard trajectories clustered in the FPCA score space.
    \end{minipage}
    \hfill
    \begin{minipage}[t]{0.49\textwidth}
        \centering
        \includegraphics[width=\textwidth]{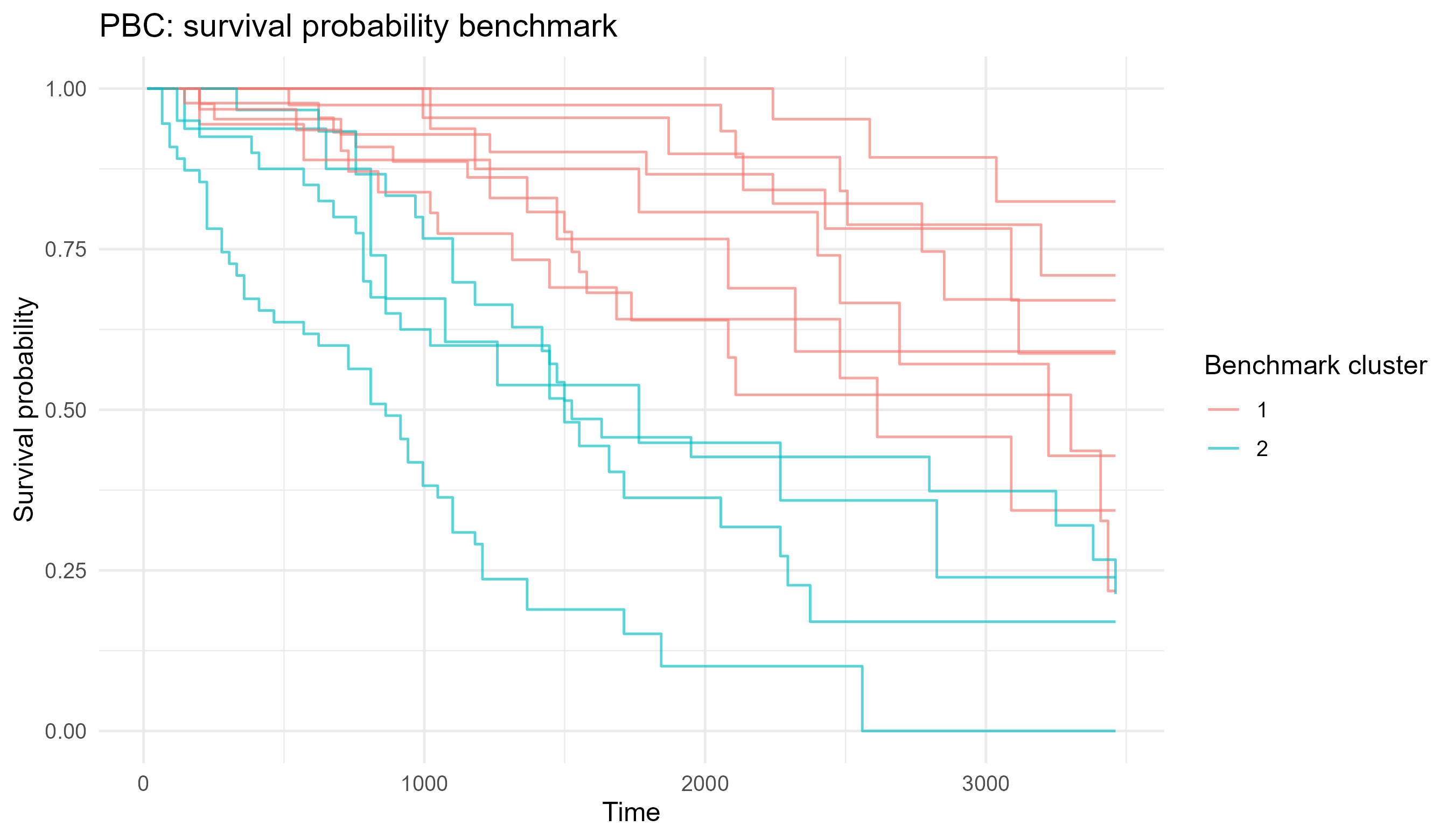}
        
        \smallskip
        \textbf{(b)} Benchmark clustering on discretized survival probability trajectories.
    \end{minipage}
    \caption{PBC clustering comparison between the proposed functional log-hazard representation and the cumulative-survival benchmark. Panel (a) reports the smoothed log-hazard trajectories clustered in the unstandardized FPCA score space. Panel (b) reports the benchmark partition obtained from discretized survival probability curves. The proposed method groups profiles according to the temporal dynamics of instantaneous risk, whereas the survival benchmark groups profiles according to cumulative survival decline.}
    \label{fig:pbc_comparison}
\end{figure*}

Figure~\ref{fig:pbc_comparison} illustrates the different information emphasized by the two representations. In the log-hazard domain, the two clusters are separated mainly by the level and temporal evolution of the instantaneous event risk. In the cumulative-survival domain, the benchmark partition reflects differences in the overall decline of survival probabilities. The two solutions are therefore not contradictory: they describe related but distinct aspects of the survival process.

The composition of the functional clusters provides an exploratory clinical interpretation. Cluster 1 contains the profiles \textit{Stg2\_BiliQ3}, \textit{Stg2\_BiliQ4}, \textit{Stg3\_BiliQ3}, \textit{Stg3\_BiliQ4}, \textit{Stg4\_BiliQ1}, \textit{Stg4\_BiliQ2}, \textit{Stg4\_BiliQ3}, and \textit{Stg4\_BiliQ4}. Cluster 2 contains the profiles \textit{Stg2\_BiliQ1}, \textit{Stg2\_BiliQ2}, \textit{Stg3\_BiliQ1}, and \textit{Stg3\_BiliQ2}. Thus, the functional partition separates profiles with higher bilirubin quartiles or Stage 4 disease from profiles corresponding to Stage 2 and Stage 3 patients with lower bilirubin quartiles. This pattern is consistent with the clinical relevance of bilirubin and disease stage, but it should be interpreted as an exploratory dynamic risk grouping rather than as a validated clinical classification.

\begin{figure*}[!t]
    \centering
    \includegraphics[width=0.85\textwidth]{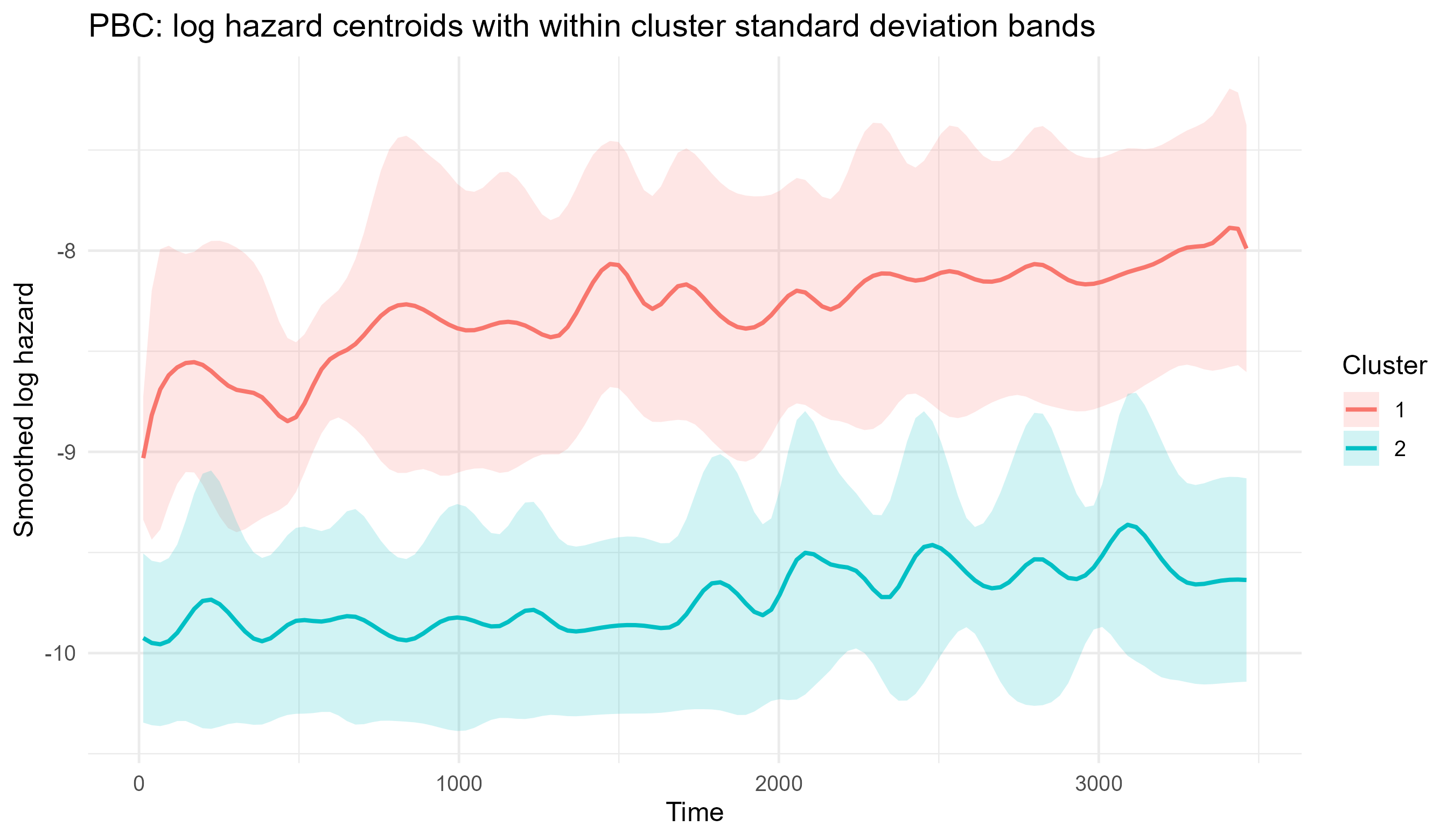}
    \caption{PBC log-hazard centroids with within-cluster standard deviation bands. The shaded regions describe functional dispersion around each centroid and should not be interpreted as confidence intervals for the mean.}
    \label{fig:pbc_centroids}
\end{figure*}

The centroid representation in Figure~\ref{fig:pbc_centroids} clarifies the temporal interpretation of the two groups. Cluster 1 exhibits a systematically higher log-hazard trajectory across most of the follow-up period, indicating higher instantaneous risk. Cluster 2 is characterized by lower log-hazard levels and a more favorable temporal profile. The separation between the two centroid curves is persistent over time, suggesting that the main contrast is not limited to a short local interval, but reflects a broad difference in the level of instantaneous event risk. The within-cluster standard deviation bands show that both clusters retain internal variability, especially in later portions of follow-up, where fewer subjects remain under observation and empirical hazard estimation is typically more variable.

\begin{figure*}[!t]
    \centering
    \begin{minipage}[t]{0.49\textwidth}
        \centering
        \includegraphics[width=\textwidth]{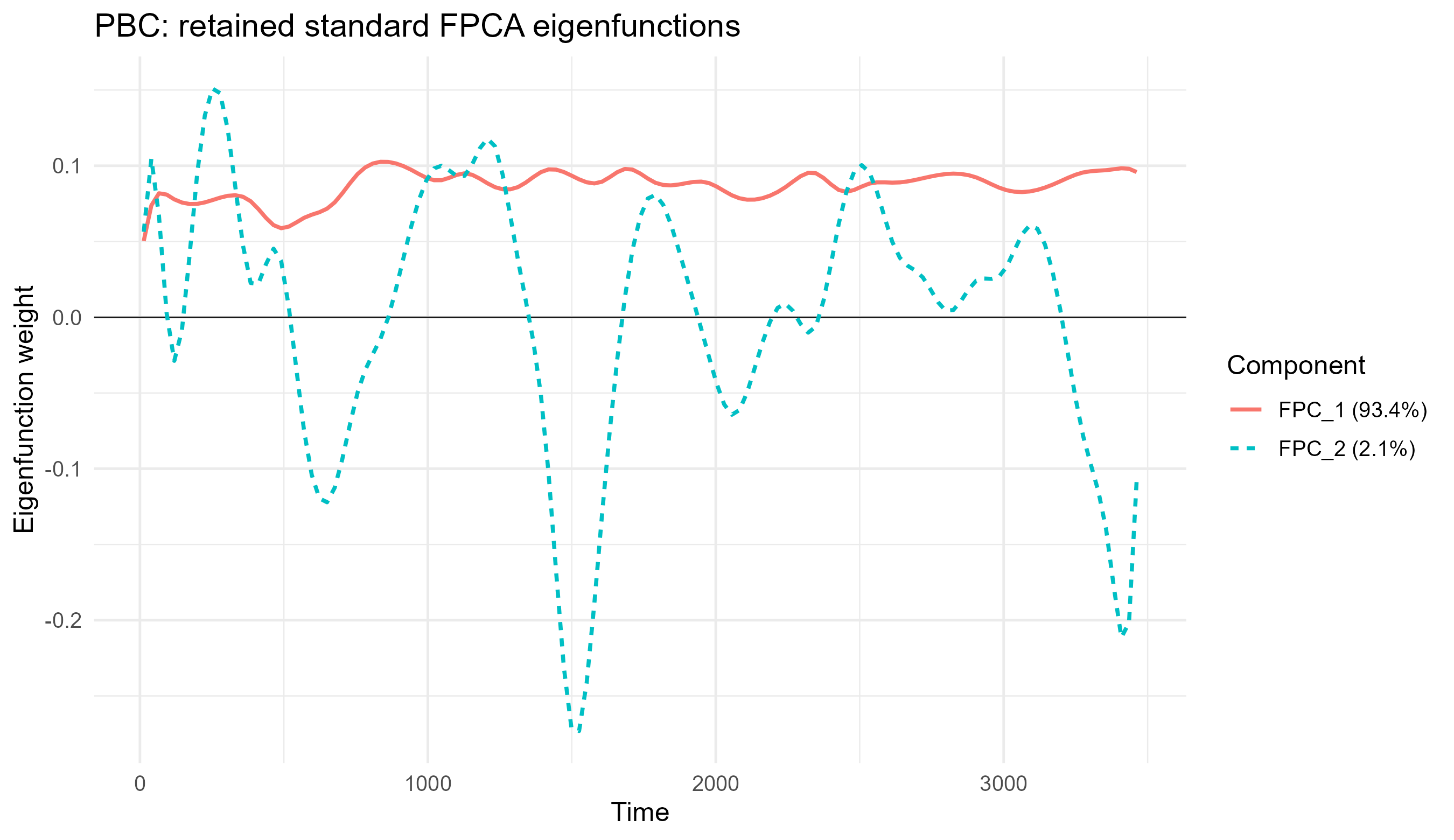}
        
        \smallskip
        \textbf{(a)} Retained standard FPCA eigenfunctions.
    \end{minipage}
    \hfill
    \begin{minipage}[t]{0.49\textwidth}
        \centering
        \includegraphics[width=\textwidth]{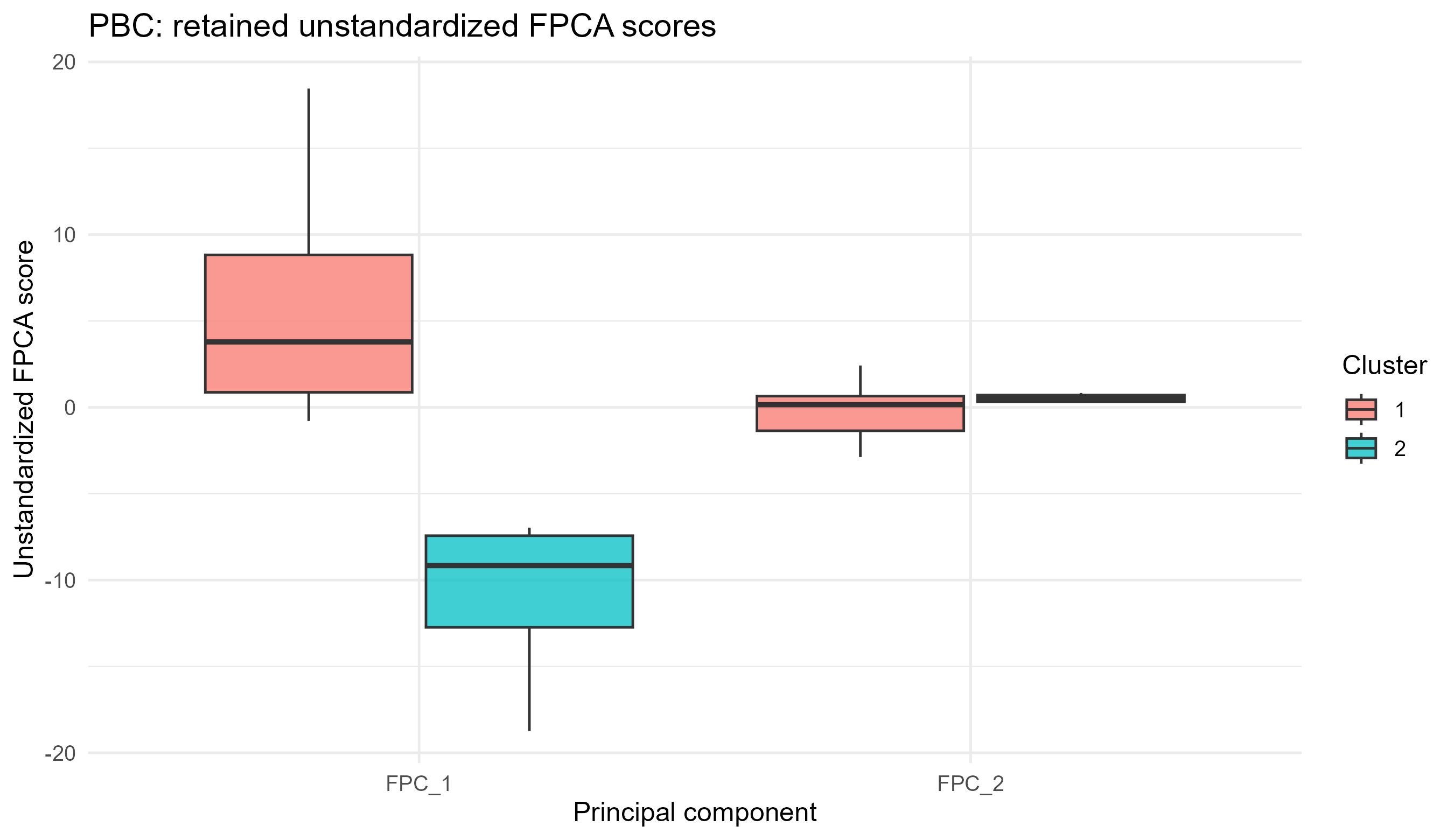}
        
        \smallskip
        \textbf{(b)} Retained unstandardized FPCA scores by cluster.
    \end{minipage}
    \caption{Post-hoc FPCA interpretation for the PBC application. Panel (a) reports the retained standard FPCA eigenfunctions extracted from the smoothed log-hazard trajectories, with the proportion of functional variance explained by each component. Panel (b) reports the distribution of the retained unstandardized FPCA scores by cluster, showing which components contribute most strongly to between-cluster separation and within-cluster heterogeneity.}
    \label{fig:pbc_fpca_explainability}
\end{figure*}

Figure~\ref{fig:pbc_fpca_explainability} provides a post-hoc interpretation of the PBC partition in the FPCA space. FPC 1 explains 93.4\% of the total functional variance and is the dominant source of separation between the two clusters. Its eigenfunction is predominantly positive over the time domain, which is consistent with an overall level effect on the log-hazard scale. The corresponding score distributions show a clear separation along FPC 1, with Cluster 1 and Cluster 2 occupying distinct score ranges. FPC 2 explains only 2.1\% of the variance and mainly captures residual temporal deviations, including local changes in the shape of the log-hazard trajectory. Therefore, the PBC clustering structure is largely driven by a dominant global risk-level contrast, with a smaller contribution from time-localized shape variation.

Bootstrap stability was used as an internal diagnostic of partition robustness conditional on the estimated functional trajectories. The mean Jaccard indices were 0.546 for Cluster 1 and 0.567 for Cluster 2, with an overall mean Jaccard value of 0.557. These values indicate moderate-to-low stability. Accordingly, the PBC partition should be interpreted with caution: it provides a coherent exploratory summary of the main log-hazard contrast in this dataset, but it does not establish a uniquely stable or externally validated clinical classification. This caution is particularly important because the analysis is based on only \(12\) aggregate trajectories, and uncertainty in the original survival and hazard estimation process is not fully propagated by bootstrap resampling of the FPCA scores alone.

\subsection{Sensitivity Analysis of the FPCA Truncation Threshold in the Real-Data Applications}

For the real-data applications, where no ground-truth partition is available, sensitivity was assessed by repeating the complete FPCA-clustering pipeline over cumulative explained-variance thresholds ranging from 90\% to 99\%. For each threshold \(c\), the number of retained FPCs was selected as
\[
M(c)
=
\min\left\{
M:
\mathrm{CPV}(M)\geq c
\right\}.
\]
The clustering procedure was then repeated in the corresponding unstandardized FPCA score space. The partition obtained under the main 95\% rule was used as the reference partition, and the partitions obtained under alternative thresholds were compared with it using the Adjusted Rand Index. In this context, ARI does not measure agreement with a true external classification, but only concordance with the reference partition obtained under the main methodological rule.

\begin{figure*}[!t]
    \centering
    \includegraphics[width=0.78\textwidth]{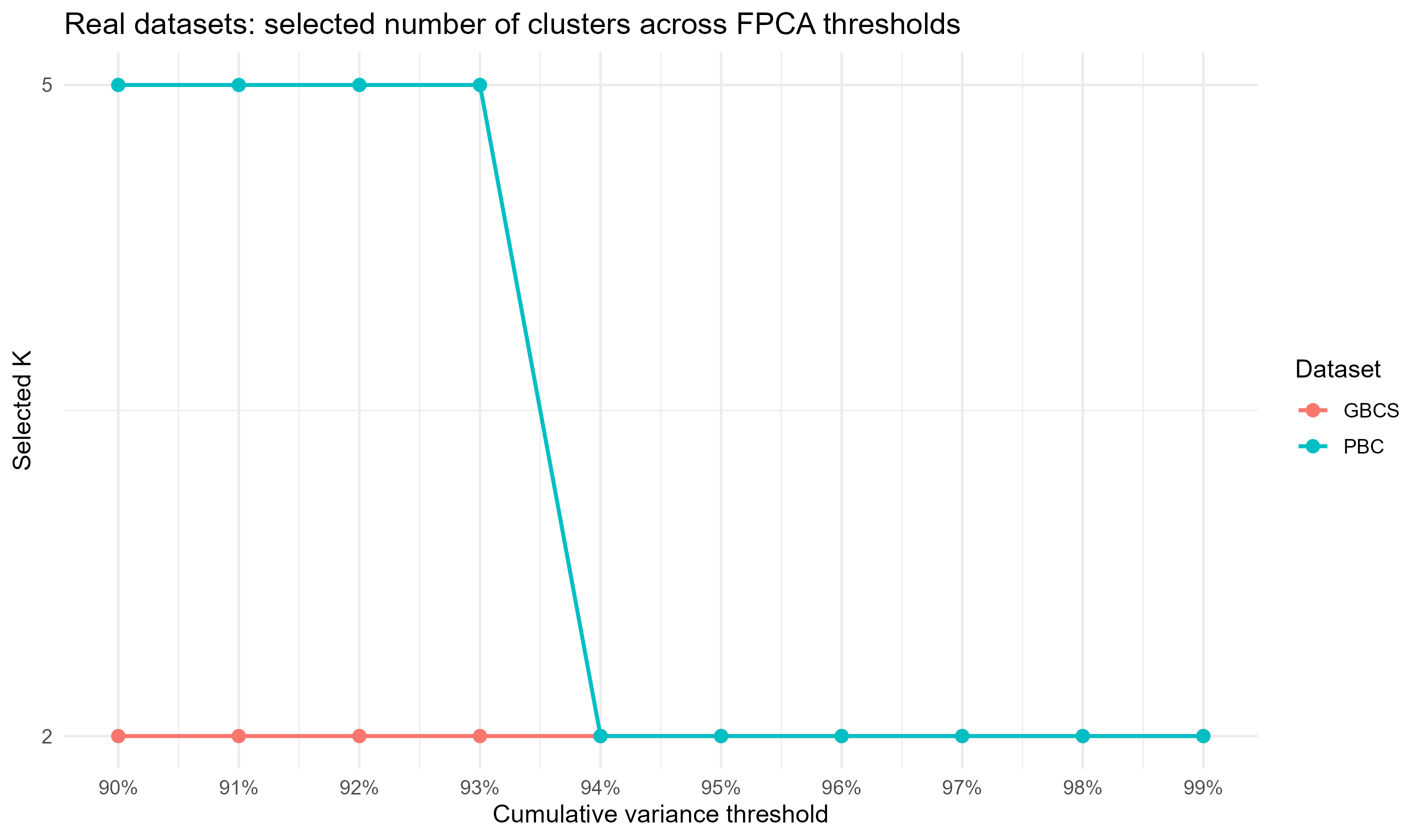}
    \caption{Selected number of clusters across FPCA truncation thresholds in the GBCS and PBC applications. The threshold ranges from 90\% to 99\% of cumulative explained functional variance.}
    \label{fig:real_threshold_sensitivity_k}
\end{figure*}

Figure~\ref{fig:real_threshold_sensitivity_k} shows distinct behaviors across the two real-data applications. In GBCS, the selected number of clusters remains constant at \(K=2\) over the entire threshold range from 90\% to 99\%. This indicates that the number of groups selected by the pipeline is robust to the amount of retained FPCA variability. In PBC, the selected number of clusters is \(K=5\) for thresholds between 90\% and 93\%, and then shifts to \(K=2\) from 94\% onward. This transition suggests that lower-dimensional representations may induce a more fragmented partition, whereas retaining a slightly larger proportion of functional variability leads to a more parsimonious and stable two-cluster structure.

\begin{figure*}[!t]
    \centering
    \includegraphics[width=0.78\textwidth]{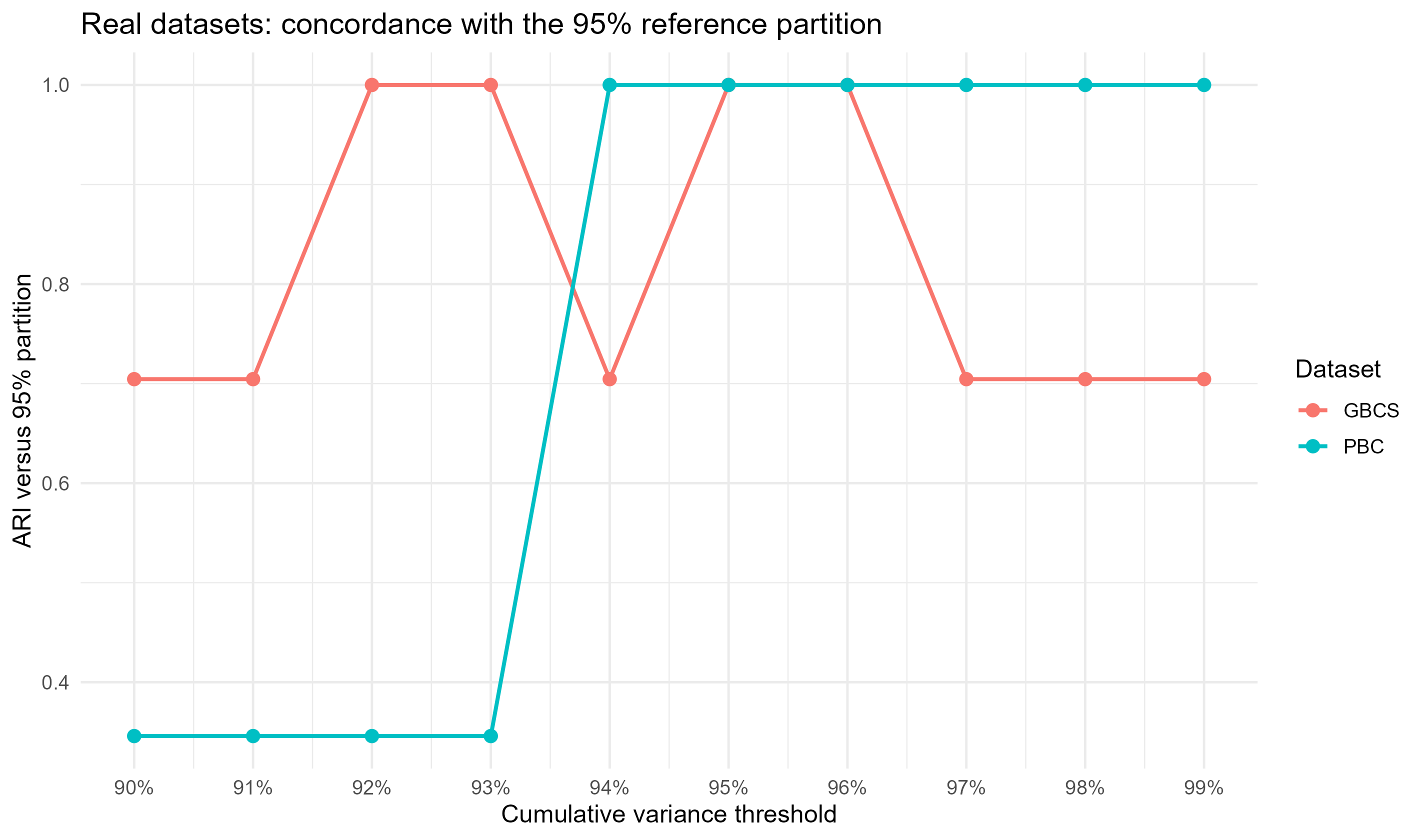}
    \caption{Concordance between alternative FPCA-threshold partitions and the 95\% reference partition in the GBCS and PBC applications. ARI is computed relative to the partition obtained under the main 95\% truncation rule.}
    \label{fig:real_threshold_sensitivity_ari}
\end{figure*}

Figure~\ref{fig:real_threshold_sensitivity_ari} complements the analysis by evaluating the concordance of each alternative threshold with the 95\% reference partition. In GBCS, the selected number of clusters remains fixed at \(K=2\), but the ARI oscillates between approximately 0.70 and 1.00. This indicates that the two-cluster structure is stable in terms of the selected \(K\), but some cohort assignments are sensitive to the precise number of retained FPCs. Therefore, the GBCS partition should be interpreted as structurally coherent but not completely invariant to FPCA truncation.

For PBC, the ARI is low for thresholds between 90\% and 93\%, consistently with the five-cluster solutions selected in that interval. From 94\% onward, the ARI becomes 1.00, indicating that all thresholds from 94\% to 99\% yield the same partition as the 95\% reference solution. This result supports the use of the 95\% rule in the PBC application: once a sufficient amount of functional variability is retained, the two-cluster structure becomes fully stable with respect to further increases in the FPCA truncation threshold.

Overall, the sensitivity analysis confirms the methodological usefulness of separating FPCA truncation from cluster validation. The 95\% rule provides a conservative functional representation, and the sensitivity analysis verifies how much the empirical partitions depend on that representation. The results are strongest for the simulation study, where recovery is invariant across thresholds. In the real-data applications, the analysis provides a more nuanced picture: PBC becomes stable once the threshold reaches 94\%, whereas GBCS shows stable selected \(K\) but partial sensitivity in cluster assignments. These findings support the use of the 95\% threshold while also indicating that real-data partitions should be interpreted as exploratory functional groupings rather than definitive classifications.

\section{Discussion and conclusions}
\label{sec:con}

This paper introduced an FDA framework for curve-level clustering in survival data, with the aim of shifting the analytical focus from cumulative survival probabilities to the temporal dynamics of instantaneous risk. The proposed approach models B-spline smoothed log-hazard trajectories as functional objects, applies standard FPCA to extract their main modes of variation, and performs clustering on the corresponding unstandardized FPCA scores. In this framework, the clustering problem is not defined on pointwise survival estimates or raw empirical hazards, but on a regularized functional representation of the log-hazard process.

The methodological rationale is that cumulative survival curves and hazard functions emphasize different aspects of the same survival process. Survival probabilities summarize the accumulated event history up to each time point, whereas hazards describe the instantaneous evolution of risk. Therefore, cohorts with similar cumulative survival profiles may still exhibit different timing, intensity, or persistence of risk peaks. By working on the log-hazard scale, the proposed method aims to capture these dynamic differences while preserving positivity of the reconstructed hazard after back-transformation.

The diagnostic comparison reported in the Appendix supports this modeling choice. The raw piecewise log-hazard estimates display substantial local instability, with abrupt jumps and isolated spikes that are especially visible in regions with sparse events or limited numbers at risk. These irregularities should not be interpreted as substantive temporal risk peaks, but as the expected instability of interval-based empirical hazard estimation. The B-spline smoothed log-hazard trajectories provide a regularized functional representation that preserves the broad temporal structure of the data while attenuating local fluctuations that are likely to reflect estimation noise. This diagnostic evidence justifies the use of FDA on smoothed log-hazard trajectories rather than on raw hazard step functions.

A central element of the final pipeline is the separation between functional representation and cluster validation. The number of retained FPCs is selected before clustering using a conservative 95\% cumulative explained-variance rule. This choice is intended to retain most of the systematic functional variability in the smoothed log-hazard trajectories, rather than to optimize the clustering result ex post. The number of clusters is then selected in the fixed FPCA score space using internal criteria and stability diagnostics. This distinction avoids conflating two different questions: how many components are needed to represent the functional trajectories, and how many groups are supported by the geometry of the represented data.

The sensitivity analyses further reinforce this distinction. In the simulation study, the recovery of the true latent groups was essentially invariant when the FPCA explained-variance threshold was varied from 90\% to 99\%. K-means on FPCA scores achieved perfect recovery across the whole threshold range, while K-medoids and Ward clustering also showed stable recovery profiles. This indicates that, in the controlled setting, the identified structure is not an artifact of the specific 95\% truncation rule.

The real-data sensitivity analyses provide a more nuanced picture. In GBCS, the selected number of clusters remained constant at \(K=2\) across the entire 90\% to 99\% threshold range, but the agreement with the 95\% reference partition varied between approximately 0.70 and 1.00. This suggests that the broad two-cluster structure is stable, while some cohort assignments remain sensitive to the precise amount of retained functional variability. In PBC, lower thresholds between 90\% and 93\% produced more fragmented five-cluster solutions, whereas thresholds from 94\% onward yielded the same two-cluster partition as the 95\% reference solution. Thus, the 95\% rule appears to be sufficiently conservative to avoid the fragmentation induced by overly compressed FPCA representations. Overall, these results support the use of the 95\% threshold while also showing that real-data partitions should be interpreted as exploratory functional structures rather than definitive classifications.

The simulation study supports the methodological construction of the proposed pipeline. Along a continuum of increasing log-hazard overlap, the recovery performance of all methods deteriorates as the latent risk regimes become less separated, as expected. However, K-means applied to the unstandardized FPCA score space showed the strongest overall recovery profile, with a more gradual loss of ARI compared with the coefficient-based benchmark and the alternative clustering procedures. The simulation case study further showed that the FPCA-based log-hazard representation can recover time-localized latent risk regimes that are less clearly identified by cumulative survival representations. These findings indicate that the benefit of the proposed approach lies in the combination of smoothing, log-hazard representation, FPCA-based dimensionality reduction, and clustering in the resulting score space.

The real-data applications provide complementary evidence. In the GBCS dataset, the proposed method selected a two-cluster partition based on seven retained FPCs explaining 96.4\% of the total functional variance. The resulting groups separated lower nodal involvement cohorts from higher nodal involvement cohorts. The first FPC explained 79.2\% of the variance and mainly represented a broad level contrast in the smoothed log-hazard trajectories. The centroid analysis confirmed that the two groups differed primarily in the overall level of instantaneous recurrence risk. However, the bootstrap stability was moderate, with an overall mean Jaccard value of 0.600. Therefore, the GBCS result should be interpreted as an internally coherent and functionally interpretable exploratory partition, not as a definitive clinical classification.

In the PBC application, the method selected a two-cluster partition based on two retained FPCs explaining 95.5\% of the total functional variance. The first FPC alone explained 93.4\%, indicating that the dominant contrast was largely a global risk-level difference on the log-hazard scale. The resulting partition separated profiles with higher bilirubin quartiles or Stage 4 disease from profiles corresponding mainly to Stage 2 and Stage 3 patients with lower bilirubin quartiles. This pattern is clinically plausible, but it must remain exploratory, especially because the analysis involved only twelve aggregate trajectories. The bootstrap stability was moderate-to-low, with an overall mean Jaccard value of 0.557, indicating that the PBC partition summarizes the main log-hazard contrast but should not be treated as a uniquely stable or externally validated classification.

Across both real-data applications, the cumulative-survival benchmark sometimes achieved comparable or higher Silhouette values than the proposed functional log-hazard method. This should not be interpreted as a contradiction or as a simple ranking of methods, because the Silhouette is computed in different representation spaces. A cohesive partition of cumulative survival curves reflects separation in accumulated survival decline, whereas a cohesive partition of smoothed log-hazard trajectories reflects separation in the temporal dynamics of instantaneous risk. The two representations therefore answer related but distinct inferential questions. The proposed method is most useful when the scientific objective is to study how risk evolves over time, rather than only how survival probability accumulates over follow-up.

The proposed framework should therefore be viewed as complementary to existing survival-curve clustering approaches. Methods based on cumulative survival probabilities are appropriate when the inferential target is the overall survival experience of groups over time. The log-hazard FDA framework is more appropriate when the target is the temporal organization of instantaneous risk, including differences in timing, persistence, and local acceleration. In applied studies, both representations may be informative, but they should not be expected to produce identical partitions because they encode different aspects of the survival process.

Several limitations should be acknowledged. First, the proposed framework operates at the curve level, using aggregate trajectories rather than individual-level covariates. This makes it appropriate for comparing groups, cohorts, centers, or clinically defined profiles, but it does not replace individual-level survival clustering or prognostic modeling. Second, the empirical hazard reconstruction and smoothing steps introduce uncertainty that is only partially captured by the subsequent bootstrap analysis on FPCA scores. A more complete uncertainty propagation would require resampling or modeling at the original individual survival-data level. Third, the real-data applications involve a limited number of aggregate trajectories, especially in the PBC case, which constrains the stability and generalizability of the resulting partitions. Fourth, although the GCV-constrained flexibility rule is designed to preserve local temporal variability while avoiding overfitting, the final clustering may still be sensitive to smoothing choices, particularly when event counts are sparse in portions of the follow-up domain. Finally, sensitivity to FPCA truncation was assessed by varying the explained-variance threshold, but other sources of sensitivity, including time-grid resolution, censoring patterns, and alternative hazard estimators, remain relevant for future investigation.

Future research may extend the present framework in several directions. One possibility is to develop fully uncertainty-aware versions of the method, in which variability from survival estimation, hazard reconstruction, smoothing, FPCA, and clustering is propagated through the entire pipeline. A second direction is the extension to multivariate or competing-risk settings, where multiple functional risk processes must be analyzed jointly. A third development concerns the integration of covariate information, either through functional regression models, supervised functional clustering, or hybrid approaches combining curve-level and individual-level information. Finally, alternative functional representations, including nonlinear functional embeddings or regularized functional autoencoders, may be explored when the geometry of the log-hazard trajectories is not adequately captured by linear FPCA.

In conclusion, the proposed framework provides a coherent and interpretable approach for clustering survival data at the curve level through smoothed log-hazard trajectories. Its main contribution is not merely to replace one clustering algorithm with another, but to redefine the object being clustered: from cumulative survival curves to functional representations of instantaneous risk dynamics. The simulation results show that this representation can improve recovery of time-localized latent risk regimes under controlled conditions, while the real-data applications illustrate how the method can produce interpretable exploratory partitions in clinical settings. The appendix diagnostics and the FPCA-threshold sensitivity analyses provide additional support for the methodological choices of the pipeline, while also clarifying the degree of uncertainty and robustness of the empirical partitions. The results support the usefulness of FDA-based log-hazard clustering as a complementary tool for survival data analysis, particularly when the timing and temporal structure of risk are central to the scientific question.

\clearpage
\appendix
\section{Diagnostic comparison between raw and smoothed log-hazard trajectories}

\begin{figure*}[!t]
\centering
\includegraphics[width=\textwidth,height=0.92\textheight,keepaspectratio]{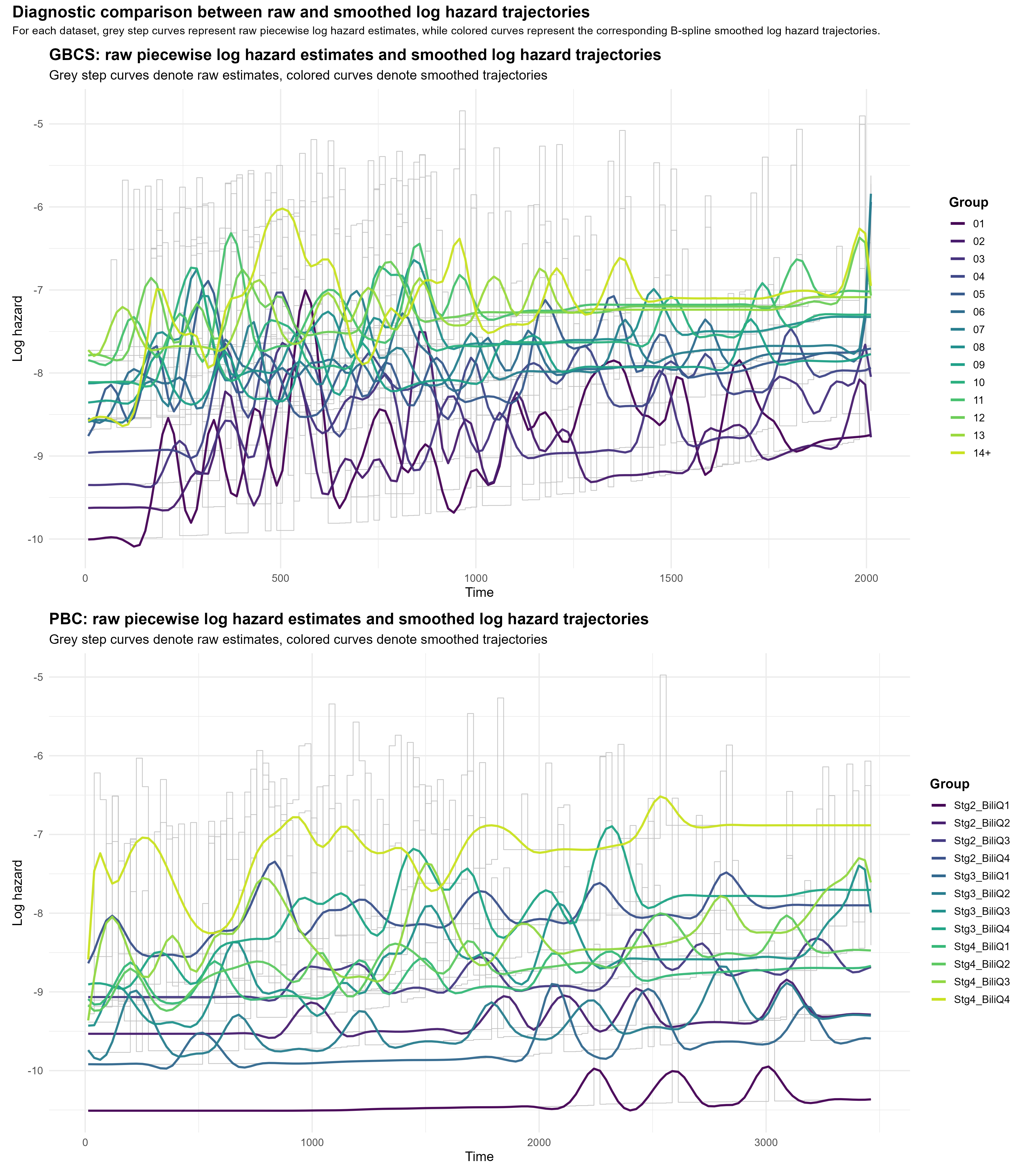}
\caption{Diagnostic comparison between raw and smoothed log-hazard trajectories for the GBCS and PBC datasets. In each panel, grey step functions represent the raw piecewise log-hazard estimates obtained prior to smoothing, whereas colored curves correspond to the B-spline smoothed log-hazard trajectories used in the functional analysis. The pronounced irregularity of the raw estimates, including local spikes and abrupt jumps, reflects the instability of interval-based hazard estimation in the presence of sparse events or small numbers at risk. The smoothed trajectories retain the main temporal patterns while attenuating unstable local fluctuations, thereby providing a more suitable functional representation for the FDA-based clustering framework.}
\label{fig:raw_smoothed_loghazard_appendix}
\end{figure*}

To document the rationale for the functional representation adopted in the main analysis, Figure~\ref{fig:raw_smoothed_loghazard_appendix} reports a diagnostic comparison between the raw piecewise log-hazard estimates and the corresponding B-spline smoothed log-hazard trajectories for the GBCS and PBC datasets.

The purpose of this diagnostic figure is methodological rather than substantive. The grey step curves represent the raw interval-based log-hazard estimates obtained before smoothing, whereas the colored curves represent the smoothed log-hazard trajectories used in the FPCA and clustering pipeline. The raw estimates are included to show the degree of local irregularity present in the empirical hazard reconstruction, not to identify clinically meaningful local peaks.

Both datasets display pronounced local instability in the raw piecewise estimates, with abrupt jumps and isolated spikes. This behavior is expected in empirical hazard estimation, especially in intervals with sparse events, limited exposure time, or a reduced number of subjects still at risk. Such fluctuations may reflect the instability of interval-based estimation rather than persistent features of the underlying risk process. For this reason, applying FPCA or clustering directly to the raw step functions would make the analysis excessively sensitive to local estimation noise and to arbitrary discretization effects.

The smoothed log-hazard trajectories provide a regularized functional representation of the same empirical information. They preserve broad temporal differences between groups while attenuating unstable local fluctuations. This step is essential for constructing a coherent functional object in \(L^2(\tau)\), for extracting interpretable principal modes of variation through FPCA, and for computing distances between trajectories in a stable score space.

The comparison also clarifies why the analysis is performed on the log-hazard scale. Working with log-hazards provides an unconstrained scale for smoothing and FPCA, while the corresponding hazard remains positive after back-transformation. Thus, the functional pipeline does not cluster raw hazard estimates, nor raw log-hazard step functions, but B-spline smoothed log-hazard trajectories.

For these reasons, the raw estimates reported in Figure~\ref{fig:raw_smoothed_loghazard_appendix} should be interpreted as diagnostic input information. The statistical analysis in the main text is based on the smoothed log-hazard trajectories, which provide the regularized functional representation required for the subsequent FPCA and clustering steps.

\section*{Declarations}

\subsection*{Acknowledgements}

The authors acknowledge the support provided within the framework of PRIN 2022 ``Spatio-temporal Functional Marked Point Processes for probabilistic forecasting of earthquake'' (Prot. 2022BN7CJP).

\subsection*{Funding}

This work was supported by PRIN 2022 ``Spatio-temporal Functional Marked Point Processes for probabilistic forecasting of earthquake'' (Prot. 2022BN7CJP).

\subsection*{Conflicts of Interest}

The authors declare that they have no financial or non-financial conflicts of interest related to the subject matter of this manuscript.

\subsection*{Data Availability Statement}

The datasets used in the real-data applications are publicly available. The German Breast Cancer Study dataset is available through the \texttt{condSURV} R package, and the Primary Biliary Cirrhosis dataset is available through the \texttt{survival} R package. The simulation study was generated using the statistical pipeline described in the manuscript.

\subsection*{Author Contributions}

Anna De Magistris, Elvira Romano, and Fabrizio Maturo jointly contributed to the conception and design of the study, methodological development, implementation of the statistical pipeline, data analysis, interpretation of the results, manuscript writing, and critical revision of the article. All authors read and approved the final version of the manuscript.

\bibliographystyle{unsrtnat}
\bibliography{wileyNJD-Chicago}


\end{document}